\documentclass[%
reprint,
superscriptaddress,
 amsmath,amssymb,
pra,
]{revtex4-1}
\usepackage[utf8]{inputenc}
\usepackage{gensymb}
\usepackage{graphicx}% Include figure files
\usepackage{dcolumn}% Align table columns on decimal point
\usepackage{bm}% bold math
\usepackage{mathtools}
\usepackage{epstopdf} % .eps figures
\usepackage{hyperref}
\usepackage{color}

\newcommand{\ket}[1]{| #1 \rangle}
\newcommand{\bra}[1]{\langle #1 |}

\newcommand{\product}[2]{\langle #1 | #2 \rangle}

\begin{document}

%\preprint{APS/123-QED}

\title{Core structure of two-dimensional Fermi gas vortices
in the BEC-BCS crossover region}

\author{Lucas Madeira}
\email{lucas.madeira@asu.edu}
\affiliation{Department of Physics, Arizona State 
University, Tempe, Arizona 85287, USA}

\author{Stefano Gandolfi}
\affiliation{Theoretical Division, Los Alamos National Laboratory,
Los Alamos, New Mexico 87545, USA}

\author{Kevin E. Schmidt}%
\affiliation{Department of Physics, Arizona State 
University, Tempe, Arizona 85287, USA}

\date{\today}% It is always \today, today,

\begin{abstract}
We report $T=0$ diffusion Monte Carlo results for 
the ground-state and vortex excitation of unpolarized spin-1/2 fermions 
in a two-dimensional disk. We investigate how vortex core structure 
properties behave over the BEC-BCS crossover. We calculate the vortex 
excitation energy, density profiles, and vortex core properties related 
to the current. We find a density suppression at the vortex core on 
the BCS side of the crossover and a depleted core on the BEC limit. 
Size-effect dependencies in the disk geometry were carefully studied.
\end{abstract}

%Fermi gas, Fermions systems (quantum statistical mechanics)

\maketitle

%\tableofcontents

\section{Introduction}
\label{sec:intro} 

The study of cold Fermi gases has proven to be a very rich research 
field, and the investigation of low-dimensional systems has become an 
active area in this context \cite{gio08,blo08}. Particularly,
the two-dimensional (2D) Fermi gas has attracted much interest 
recently. It was 
the object of several theoretical investigations 
\cite{ran89,ran90,pet03,mar05,tem07,zha08}, but its experimental 
realization, using a highly anisotropic potential, was a milestone in 
the study of these systems \cite{kir10}. Many other studies have been 
carried out since \cite{ore11,mak14}. Quantum Monte Carlo (QMC) methods 
were successfully employed to compute several properties of the BEC-BCS 
crossover.
These methods include diffusion Monte Carlo (DMC) \cite{ber11,gal16}, 
auxiliary-field quantum Monte Carlo \cite{shi15}, and lattice 
Monte Carlo \cite{and15,ram16,luo16}. The fact that a fully attractive 
potential in 
2D always supports a bound state, and the ability to vary the 
interaction strength over the entire BEC-BCS crossover regime offers 
rich 
possibilities for the study of these systems.
	
The presence of quantized vortices is an indication of a superfluid 
state in both Bose and Fermi systems. In three-dimensional (3D) systems, much progress has been 
made \cite{bul03,sen06,sim13,mad16}, including the observation of vortex 
lattices in a strongly interacting rotating Fermi gas of $^6$Li 
\cite{zwi05}. With the recent progress on the 2D Fermi gases, it seems 
natural to also extend the theoretical study of vortices to these 
systems. Interest 
is further augmented in 2D, where a Berezinksii-Kosterlitz-Thouless 
transition \cite{ber70,kos72} could take place at finite temperatures, 
and pairs of vortices and antivortices would eventually condense to form 
a 
square lattice \cite{bot06}.

We are interested in how the properties of a vortex change over the 
BEC-BCS crossover. In this work we focus on ultracold atomic Fermi 
gases, but 
it is noteworthy that a duality is expected between neutron matter and 
superfluid atomic Fermi gases. In 3D, both ultracold atomic gases and
low-density neutron matter exhibit pairing gaps of the order of the 
Fermi 
energy \cite{bro13}. Neutron-matter properties depend on the 
interaction strength and, unlike the Fermi atom gases, the possibility 
of microscopically tuning interactions of neutron-matter is not 
available. However, we can study neutron pairing by looking at the BCS 
side of the crossover \cite{Gezerlis:2008,gez10}. Vortex properties are 
also 
of significant interest in neutron matter \cite{bla99,yu03}
because a significant part of the matter in rotating neutron 
stars is superfluid, and vortices are expected to appear. Moreover, 
phases 
called nuclear pasta, where neutrons are restricted to 1D or 2D 
configurations, are predicted in neutron stars \cite{rav83,yu03}.

We report properties of a single vortex in a 2D Fermi gas. We 
considered the ground-state to be a disk with hard walls and total 
angular momentum zero, and the vortex excitation corresponds to each 
fermion pair having angular momentum $\hbar$. Hopefully, our results 
will motivate experiments to increase our understanding of vortices in 
2D Fermi gases.

This work is structured as it follows. In Sec.~\ref{sec:met} we 
introduce the methodology employed. In Sec.~\ref{sec:finite} we discuss 
aspects of finite-size fermionic systems, we briefly introduce 2D 
scattering in Sec.~\ref{sec:scatt}, Sec.~\ref{sec:wf} is devoted to the 
wave functions employed for the bulk, disk, and vortex systems, and
we summarize the employed QMC methods in Sec.~\ref{sec:qmc}. The 
results are presented in Sec.~\ref{sec:res}. Sec.~\ref{sec:disk} 
contains the ground-state energies in the disk geometry and discussions 
on size-effects. In Sec.~\ref{sec:vortex} we present the vortex 
excitation energy. The determination of the crossover region is done in 
Sec.~\ref{sec:chempot}. Density profiles of the vortex and ground-state 
systems are shown in Sec.~\ref{sec:density}. Properties of the vortex 
core are discussed in Sec.~\ref{sec:current}. Finally, a summary of the 
work is presented in Sec.~\ref{sec:sum}.

\section{Methods}
\label{sec:met}

Previous simulations of vortices in 3D bosonic systems, such as $^4$He, 
have often employed a periodic array of counter-rotating vortices, 
which enables the usage of periodic boundary conditions. In the $^4$He 
calculations of Ref.~\onlinecite{sad97}, the simulation cell consisted 
of 300 particles in four counter-rotating vortices. If we had employed 
a similar methodology, we would need the same number of fermion pairs, 
i.e., a system with 600 fermions. There are simulations of 
fermionic systems that have been performed with this number of 
particles, but the variance required for a detailed optimization is 
beyond the scope of this work. Instead, we considered a disk geometry  
similar to the one used in Ref.~\cite{Ortiz:1995} for DMC simulations of 
the vortex core structure 
properties in $^4$He.

\subsection{Finite-size systems}
\label{sec:finite}

We are interested in the interacting many-body problem, but it is 
useful to first consider the non interacting case. In this section we 
compare the energy of finite-size 2D systems to the results in the 
thermodynamic limit.

First let us consider the case of $N$ fermions in a square of side $L$ 
with periodic boundary conditions. The single-particle states are plane 
waves $\psi_{\textbf{k}_{\textbf{n}}}
(\textbf{r})=e^{i\textbf{k}_\textbf{n}\cdot \textbf{r}}/L$, with wave 
vector
\begin{eqnarray}
\label{eq:kbulk}
\textbf{k}_{\textbf{n}}=\frac{2\pi}{L}(n_x 
\hat{\textbf{x}}+n_y\hat{\textbf{y}}).
\end{eqnarray}
The eigenenergies are $E_\textbf{n}=\hbar^2 
\textbf{k}_{\textbf{n}}^2/2m$, where $m$ is the mass of the fermion. At 
$T=0$, all states with energy up to the Fermi energy 
$\epsilon_F=\hbar^2k_F^2/2m$, where $k_F$ is the Fermi wave number, are 
occupied. A shell structure arises from the fact that different 
combinations of $n_x$ and $n_y$ in Eq.~(\ref{eq:kbulk}) yield the same 
$|\textbf{k}_\textbf{n}|$. The closed shells occur at total particle 
number $N=(2,10,18,26,42,50,58,\cdots)$. The free gas energy of a 
finite system with $N$ fermions, $E_{FG}^{\rm bulk}(N)$, is readily 
calculated by 
filling the lowest energy states described by Eq.~(\ref{eq:kbulk}). In 
the thermodynamic limit, which corresponds to $N,L\to \infty$ and 
$n=N/L^2$ held 
constant, the energy per particle of the free gas is $E_{\rm 
FG}=\epsilon_F/2$ and $k_F=\sqrt{2\pi n}$.

Now let us consider the case of $N$ fermions in a disk of radius 
$\mathcal{R}$ with a hard wall boundary condition, i.e., the wave 
function must vanish at $\mathcal{R}$. The single-particle states are
\begin{eqnarray}
\label{eq:spdisk}
\psi_{\nu p}(\rho,\varphi)=\mathcal{N}_{\nu p} J_\nu\left(\frac{j_{\nu 
p}}{\mathcal{R}}\rho\right) e^{i \nu \varphi},
\end{eqnarray}
where $(\rho,\varphi)$ are the usual polar coordinates, 
$\mathcal{N}_{\nu p}$ is a normalization constant, $J_\nu$ are Bessel 
functions of the first kind, and $j_{\nu p}$ is the $p$-th zero of 
$J_\nu$. The quantum number $\nu$ can take the values $0,\pm 1, \pm 2, 
\cdots$ and $p=1,2,\cdots$. The corresponding eigenenergies are
\begin{eqnarray}
\label{eq:enup}
E_{\nu p}=\frac{\hbar^2}{2m}\left(\frac{j_{\nu p}}
{\mathcal{R}}\right)^2.
\end{eqnarray}
This system also presents a shell structure, due to the energy 
degeneracy of single-particles states with the same $|\nu|$, with shell 
closures at total particle number $N=(2,6,10,12,16,20,24,28,30,34,
\cdots)$. Notice that the energy levels of the bulk system are much 
more degenerate than the ones of the disk. In practice this means that 
more shells are needed to describe a disk with a given $N$. The free gas 
energy for the disk, $E^{\rm disk}_{FG}(N)$, can be calculated 
analogously to 
the bulk case using the energy levels of Eq.~(\ref{eq:enup}). The 
thermodynamic limit 
for this case corresponds to ${\cal R}\to \infty$ with $n=N/(\pi {\cal 
R}^2)$ held constant, and $E_{FG}$ and $k_F$ go to the same expressions 
as the bulk ones.

\begin{figure}[!htb]
  \centering
  \includegraphics[angle=-90,width=\linewidth]{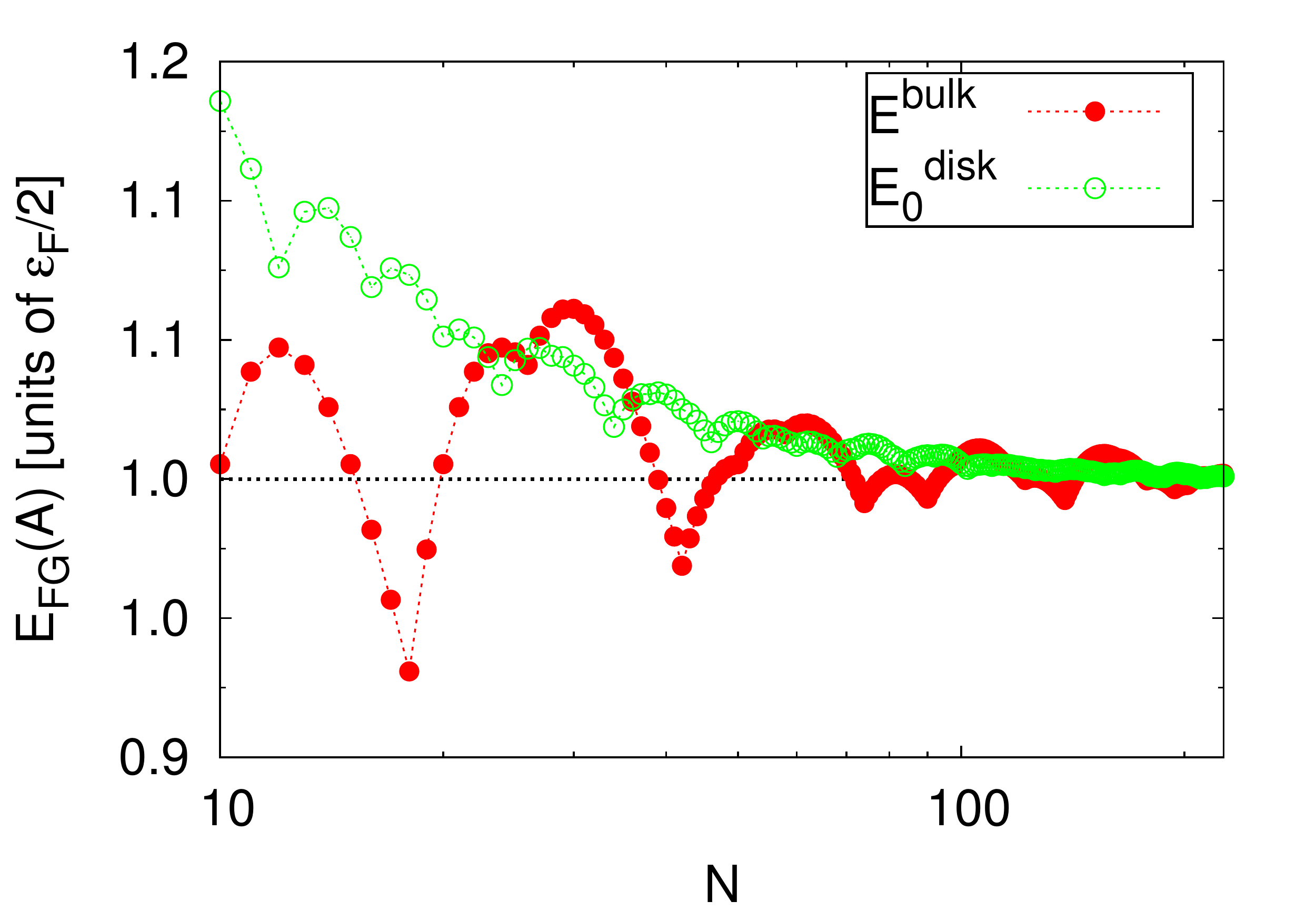}
  \caption{(Color online) Free-gas energy for finite-size systems as a 
function of the number of particles $N$, where the dotted lines are 
drawn to guide the eye.
The (red) closed circles denote the energy of the 
bulk system,
$E_{FG}^{\rm bulk}(N)$,
and the (green) open circles indicate the bulk energy component in the 
disk 
geometry, $E_0^{\rm disk}(N)$, as defined in Eq.~(\ref{eq:diskbulk}). 
Local minima in $E_{FG}^{\rm bulk}(N)$ correspond to shell closures.}
  \label{fig:efg}
\end{figure}

The comparison between the free gas energy of finite systems in the bulk 
case and in the disk geometry is not immediate due to the presence of 
hard walls in the latter.
In order to compare the free gas energy in both geometries, we define  
\begin{eqnarray}
\label{eq:diskbulk}
E_0^{\rm disk}(N)=
E_{FG}^{\rm disk}(N)-\frac{\lambda_s}{2}\sqrt{\frac{n}{\pi N}},
\end{eqnarray}
in which we separated the total energy $E_{FG}^{\rm disk}(N)$ into a 
bulk component, $E_0^{\rm disk}(N)$, and a surface term, the second term 
on the RHS. For further discussions on the functional form of the 
surface term, see Sec.~\ref{sec:disk}.
Figure~\ref{fig:efg} shows $E_{FG}^{\rm bulk}(N)$ and $E_0^{\rm disk}(N)$, 
with $\lambda_s=17.5 \ E_{FG} k_F^{-1}$, at the same density.
The value of $\lambda_s$, within a 0.2\% error, was determined by  
fitting the data for
$10 \leqslant N \leqslant 226$ to the functional form of 
Eq.~(\ref{eq:diskbulk}).

The disk presents a considerably higher 
free gas energy, if compared to the bulk system, due to the presence of 
hard walls, but the difference between them is rapidly suppressed as we 
increase the particle number.

\subsection{Scattering in 2D}
\label{sec:scatt}

Two-body scattering by a finite-range potential $V(r)$ in 2D is 
described by the Schrödinger equation. We separate the solutions into 
radial $R(r)$ and angular $P(\phi)$ parts, the latter being a constant 
for 
$s$-wave scattering.
The two-body equation for an azimuthally symmetric ($s$-wave) solution 
is
\begin{eqnarray}
\label{eq:scatt}
\left[ -\frac{\hbar^2 \nabla^2}{2m_r}+V(r)\right]u(r)=
\frac{\hbar^2 k^2}{2m_r}u(r),
\end{eqnarray}
where $m_r$ is the reduced mass of the system, and $\hbar^2k^2/2m_r$ is 
the scattering energy. The scattering length $a$ and effective range 
$r_{\rm eff}$ can be easily determined from the $k\to 0$ solution of 
Eq.~(\ref{eq:scatt}), $u_0(r)$, and its asymptotic form $y_0$. We 
choose the solution
\begin{eqnarray}
y_0(r)=-\ln\left(\frac{r}{a}\right),
\end{eqnarray}
and we match $u_0$ and $y_0$, and their derivatives, outside the range 
of the potential.

In 2D, the low-energy phase shifts $\delta(k)$, $a$, and effective range 
$r_{\rm eff}$, 
are related by \cite{khu09}
\begin{equation}
\label{eq:shape_ind}
\cot \delta(k) \approx \frac{2}{\pi}\left[\ln\left( \frac{k a}{2} 
\right) + \gamma \right] + \frac{k^2 r_{\rm eff}^2}{4},
\end{equation} 
where $\gamma=0.577\dots$ is the Euler-Mascheroni constant, and the 
effective range is defined as \cite{adh86}
\begin{eqnarray}
r_{\rm eff}^2 = 4 \int_0^\infty (y_0^2(r)-u_0^2(r))r\ dr.
\end{eqnarray}
Equation~(\ref{eq:shape_ind}) is often called the shape-independent 
approximation because it guarantees that a broad range of well-chosen 
potentials can be constructed to describe low-energy scattering. We 
consider the modified Poschl-Teller potential
\begin{eqnarray}
\label{eq:poschl}
V(r)= - v_0 \frac{\hbar^2}{m_r} \frac{\mu^2}{\cosh^2(\mu r)},
\end{eqnarray}
where $v_0$ and $\mu$ can be tuned to reproduce the desired $a$ and 
$r_{\rm eff}$.

Bound-states occur for purely attractive potentials for any strength in 
2D. If we continually increase the depth of $V(r)$, $a$ will eventually 
reach zero, and then it diverges to $+ \infty$ when a new bound-state is 
created. The binding energy of the pair is given by
\begin{eqnarray}
\label{eq:eb}
\epsilon_b = -\frac{4\hbar^2}{ma^2 e^{2\gamma}}.
\end{eqnarray}
We chose values of $v_0$ and $\mu$ such that only one bound-state is 
present, and $k_F r_{\rm eff}$ is held constant at 0.006 \cite{gal16}.
This choice guarantees that the systems studied in this work are in the 
dilute regime, since $r_0 \gg r_{\rm eff}$, where $r_0=1/\sqrt{\pi n}$ 
is of order of the interparticle spacing.

\subsection{Wave functions}
\label{sec:wf}

The BCS wave function, which describes pairing explicitly, has been 
successfully used in a variety of strongly interacting Fermi gases 
systems, such as: 3D \cite{car03} and 2D \cite{gal16} bulk systems, 
vortices in the unitary regime \cite{mad16}, two-component mixtures 
\cite{gez09,gan14}, and many other systems.
This wave function, projected to a fixed number
of particles $N$ (half with spin-up and half with spin-down),
can be written as the antisymmetrized product \cite{bou88}
\begin{flalign}
\label{eq:anti_all}
\psi_{\rm BCS}(\textbf{R},S) = \mathcal{A}[ \phi(\textbf{r}_1,s_1, 
\textbf{r}_{2},s_{2}) \phi(\textbf{r}_3,s_3,
\textbf{r}_{4},s_{4})\dots \nonumber \\
\phi(\textbf{r}_{N-1},s_{N-1},
\textbf{r}_{N},s_{N})],
\end{flalign}
where $\textbf{R}$ is a vector containing the particle positions
$\textbf{r}_i$, $S$ stands for the spins $s_i$, and $\phi$ is the
pairing function, which is given by
\begin{flalign}
\phi(\textbf{r},s,\textbf{r}',s') = \tilde{\phi}(\textbf{r},
\textbf{r}') \left[ \product{s \ s'}{\uparrow \ \downarrow} - 
\product{s \ s'}{ \downarrow \ \uparrow} \right],
\end{flalign}
where we have explicitly included the spin part to impose singlet 
pairing. The assumed expressions for $\tilde{\phi}$ depend on the 
system being studied (see Secs. \ref{sec:wfbulk}, \ref{sec:wfdisk}, and 
\ref{sef:wfvortex}).
Since neither the Hamiltonian or any operators in the quantities we 
calculate flip the spins, we adopt hereafter the convention of primed 
indexes 
to denote spin-down particles and
unprimed ones to refer to spin-up particles. Equation~(\ref{eq:anti_all}) 
reduces to
\begin{flalign}
\label{eq:BCS_anti}
&\psi_{\rm BCS}(\textbf{R},S) = \mathcal{A}[ \phi(\textbf{r}_1,s_1, 
\textbf{r}_{1'},s_{1'}) \nonumber \\ 
&\phi(\textbf{r}_2,s_2,
\textbf{r}_{2'},s_{2'})\dots
\phi(\textbf{r}_{N/2},s_{N/2},\textbf{r}_{N/2'},s_{N/2'})],
\end{flalign}
where the antisymmetrization is over spin-up and/or spin-down particles 
only.
This wave function can be 
calculated efficiently as
a determinant \cite{gan09}.

In addition to fully paired systems, it is also possible to simulate 
systems with unpaired particles \cite{car03}, described by single 
particle states $\Phi(\textbf{r})$. For $q$ pairs, $u$ spin-up, and $d$ 
spin-down unpaired single particles states, $N=2q+u+d$, we can rewrite 
Eq.~(\ref{eq:BCS_anti}) as
\begin{flalign}
\label{eq:unpaired}
\psi_{\rm BCS}(\textbf{R},S) = \mathcal{A}[
\phi(\textbf{r}_1,s_1, \textbf{r}_{1'},s_{1'}) \cdots \nonumber \\
\phi(\textbf{r}_q,s_q, \textbf{r}_{q'},s_{q'})
\Phi_{1\uparrow}  (\textbf{r}_{q+1})\cdots
\Phi_{u\uparrow}  (\textbf{r}_{q+u}) \nonumber \\
\Phi_{1\downarrow}(\textbf{r}_{(q+1)'})\cdots
\Phi_{d\downarrow}(\textbf{r}_{(q+d)'})].
\end{flalign}

We also included a two-body
Jastrow factor $f(r_{ij'})$, $r_{ij'}=|\textbf{r}_i-\textbf{r}_{j'}|$, 
which accounts for
correlations between antiparallel spins. It is obtained from solutions 
of 
the two-body
Schrödinger's equation
\begin{equation}
\left[ -\frac{\hbar^2 \nabla^2}{2m_r} + V(r) 
\right]f(r<d)=\lambda f(r<d),
\end{equation}
with the boundary conditions $f(r>d)=1$ and $f'(r=d)=0$,
where $d$ is a variational parameter, and $\lambda$ is adjusted so that 
$f(r)$ is nodeless. The total trial wave function is written as
\begin{equation}
\psi_{\rm T}(\textbf{R},S)=\prod_{i,j'} f(r_{ij'}) \psi_{\rm BCS}
(\textbf{R},S).
\end{equation}

\subsubsection{Bulk system}
\label{sec:wfbulk}

The assumed form of the pairing function for the bulk case is the same 
as Ref.~\cite{car03},
\begin{flalign}
\label{eq:bulk}
\tilde{\phi}_{\rm bulk}(\textbf{r},\textbf{r}') = \sum_{n=1}^{n_c} 
\alpha_n e^{i \textbf{k}_{\textbf{n}} \cdot (\textbf{r} - \textbf{r}')} 
+ \tilde{\beta}(|\textbf{r}-\textbf{r}'|),
\end{flalign}
where $\alpha_n$ are variational parameters, and contributions from 
momentum states up to a level $n_c$ are included. Contributions with 
$n>n_c$ are included through the $\tilde{\beta}$ function given by
\begin{equation}
\tilde{\beta}(r)=
\begin{cases}
\beta(r)+\beta(L-r)-2\beta(L/2)
 &\text{for } r\leqslant L/2\\
0 &\text{for } r > L/2
\end{cases}
\end{equation}
with
\begin{eqnarray}
\label{eq:beta}
\beta(r) = [1+c b r][1-e^{-dbr}]\frac{e^{-br}}{dbr},
\end{eqnarray}
where $r=|\textbf{r}-\textbf{r}'|$ and $b$, $c$, and $d$ are
variational parameters.
This functional form of $\beta(r)$ describes the short-distance
correlation of particles with antiparallel spins.
We consider $b$ = 0.5 $k_F$, $d=5$, and $c$ is adjusted so that 
$\partial \tilde{\beta}/\partial r=0$ at $r=0$.

\subsubsection{Disk}
\label{sec:wfdisk}

The pairing function for the disk geometry is constructed using the 
single-particle orbitals of Eq.~(\ref{eq:spdisk}). Each pair consists 
of one single-particle orbital coupled with its time-reversed state. 
This ansatz has been used before in the 3D system \cite{mad16}, a 
cylinder with hard walls, and the form presented here is analogous to 
that one if we disregard the $z$ components. We supposed the pairing 
function to be 
\begin{flalign}
\label{eq:disk}
\tilde{\phi}_{\rm disk}(\textbf{r},\textbf{r}') = \sum_{n=1}^{n_c} 
\tilde{\alpha}_n
\mathcal{N}_{\nu p}^2 J_\nu\left(\frac{j_{\nu p}}
{\mathcal{R}}\rho\right) J_\nu\left(\frac{j_{\nu p}}{\mathcal{R}}\rho' 
\right) e^{i \nu (\varphi-\varphi')} \nonumber \\ + 
\bar{\beta}(\textbf{r},
\textbf{r}'),
\end{flalign}
where
the $\tilde{\alpha}_n$ are variational parameters, and
$n$ is a label for the disk shells, such that different states with the 
same energy are associated with the same variational parameter.
The 
$\bar{\beta}$ function is similar to $\tilde{\beta}$ employed in the 
bulk system, but we modify it to ensure the hard wall boundary 
condition is met,
\begin{equation}
\label{eq:betadisk}
\bar{\beta}(\textbf{r},\textbf{r}')=
\begin{cases}
{\cal N}_{01}^2 J_0\left(\frac{j_{01}\rho}{{\cal R}}\right) 
J_0\left(\frac{j_{01}\rho'}{{\cal R}}\right)\times \\
\left[ 
\beta(r)+\beta(2{\cal R}-r)-2\beta({\cal R})\right]
 &\text{for } r\leqslant {\cal R}\\
0 &\text{for } r > {\cal R}
\end{cases}
\end{equation}
and $\beta$ has the same expression as the bulk case, 
Eq.~(\ref{eq:beta}).

\subsubsection{Vortex}
\label{sef:wfvortex}

The vortex excitation is accomplished by considering pairing orbitals 
which are eigenstates of $L_z$ with eigenvalues $\pm \hbar$. This is 
achieved by coupling single-particle states with angular quantum 
numbers differing by one. In this case we used pairing orbitals of the 
form
\begin{flalign}
\label{eq:vortex}
&\tilde{\phi}_{\rm vortex}(\textbf{r},\textbf{r}') = \sum_{n=1}^{n_c} 
\bar{\alpha}_n
\mathcal{N}_{\nu p} \mathcal{N}_{\nu-1 ; p}\times \nonumber \\
&\left\{ J_\nu\left(\frac{j_{\nu p}}{\mathcal{R}}\rho\right) 
J_{\nu-1}\left(\frac{j_{\nu-1;p}}{\mathcal{R}}\rho' \right) e^{i (\nu 
\varphi- (\nu-1) \varphi')} \right. \nonumber \\
&+ \left. J_\nu\left(\frac{j_{\nu p}}{\mathcal{R}}\rho'\right) 
J_{\nu-1}\left(\frac{j_{\nu-1;p}}{\mathcal{R}}\rho \right) e^{i (\nu 
\varphi'- (\nu-1) \varphi)}\right\},
\end{flalign}
where $n$ is a label for the vortex shells, and $\bar{\alpha}$ are 
variational parameters.
The largest contribution is assumed to be from states with the same 
quantum number $p$ for the radial part \cite{mad16}. 
Equation~(\ref{eq:vortex}) is symmetric under interchange of the prime and 
unprimed coordinates, as required for singlet 
pairing.

The $\bar{\beta}$ function of Eq.~(\ref{eq:betadisk}) is not suited to 
describe the vortex state because it is an eigenstate of $L_z$ with angular 
momentum zero. We tried different functional forms that had the desired 
angular momentum eigenvalue, but none of them resulted in a significant lower 
total energy. Thus, we chose to employ only the terms in 
Eq.~(\ref{eq:vortex}).

\subsection{Quantum Monte Carlo}
\label{sec:qmc}

The Hamiltonian of the two-component Fermi gas is given by
\begin{equation}
H=-\frac{\hbar^2}{2m}\left[ \sum_{i=1}^{N_\uparrow} \nabla_i^2 + 
\sum_{i=j'}^{N_\downarrow} \nabla_{j'}^2 \right] + \sum_{i,j'} 
V(r_{ij'}),
\end{equation}
with $N=N_\uparrow+N_\downarrow$, and $V(r_{ij'})$ given by 
Eq.~(\ref{eq:poschl}).
The DMC method projects the lowest energy state of $H$ from an initial 
state $\psi_T$, obtained from variational Monte Carlo (VMC) 
simulations. The propagation, which is carried out in imaginary time 
$\tau$, can be written as
\begin{equation}
\psi(\tau)= e^{-(H-E_T)\tau}\psi_T,
\end{equation}
where $E_T$ is an energy offset. In the $\tau\to\infty$ limit, only the 
lowest energy component $\Phi_0$ survives
\begin{equation}
\lim_{\tau \to \infty} \psi(\tau)=\Phi_0.
\end{equation}
The imaginary time evolution is given by
\begin{equation}
\label{eq:green}
\psi(\textbf{R},\tau)=\int d\textbf{R}' G(\textbf{R},\textbf{R}',
\tau)\psi_T(\textbf{R}'),
\end{equation}
where $G(\textbf{R},\textbf{R}',\tau)$ is the Green's function 
associated with $H$. The Green's function contains two pieces, a 
diffusion term related to the kinetic operator, and a branching term 
related to the potential. We solve an importance sampled version of 
Eq.~(\ref{eq:green}) iteratively, using the Trotter-Suzuki 
approximation to evaluate $G(\textbf{R},\textbf{R}',\tau)$, which 
requires the time steps $\Delta \tau$ to be small. We circumvent the 
fermion-sign problem by using the fixed-node approximation, which 
restricts transitions across a chosen nodal surface \cite{fou01}. Hence 
our estimates of energy expectation values are upper bounds.

We carefully optimized the trial wave function $\psi_T$, since it is 
used in three ways: an approximation of the ground-state in the VMC 
calculations, as an importance function, and to give the nodal surface 
for the fixed-node 
approximation. The variational parameters
\footnote{See Supplemental Material at [\textit{url to be inserted by 
publisher}] for the variational parameters of Eqs.~(\ref{eq:bulk}), 
(\ref{eq:disk}), and (\ref{eq:vortex})}
in Eqs.~(\ref{eq:bulk}), (\ref{eq:disk}), and (\ref{eq:vortex}) were 
determined using the stochastic reconfiguration method \cite{cas04}.

Expectation values of operators that do not commute with the 
Hamiltonian, for example the current and density, were calculated using 
extrapolated estimators \cite{cep86}
\begin{equation}
\label{eq:ext}
\bra{\Phi} \hat{S} \ket{\Phi} \approx 2 \bra{\Phi} \hat{S} \ket{\psi_T} 
- \bra{\psi_T} \hat{S} \ket{\psi_T} + {\cal O}[(\Psi-\psi_T)^2],
\end{equation}
where we combine the results of VMC and DMC runs.

\section{Results}
\label{sec:res}

We define the interaction strength $\eta \equiv \ln(k_F a)$. Large 
values of $\eta$ correspond to the BCS side of the crossover, while 
small $\eta$ are on the BEC side. We probed $0.0 \leqslant \eta 
\leqslant 1.5$, which encompasses the crossover region (see 
Sec.~\ref{sec:chempot}). For all systems the number density is
$n=N/(\pi {\cal R}^2)$, and $k_F=\sqrt{2N}/{\cal R}$.

\subsection{Ground-state energy and size-effects}
\label{sec:disk}

We used the pairing function of Eq.~(\ref{eq:bulk}), and $N=26$, to 
calculate 
the ground-state energy per particle of the bulk systems. 
Our results (see Table \ref{tab:bulk}) are in agreement with previous DMC 
calculations \cite{gal16}.

\begin{table}[htbp]
\caption{Comparison between the ground-state energy per particle of the 
bulk ($E_{\rm bulk}$) and disk systems as a function of the interaction 
strength $\eta$. The parameters $E_0$ and $\lambda_s$, see 
Eq.~(\ref{eq:eofr}), are related to our assumption of the functional 
form of the ground-state energy per particle in the disk geometry. 
}
\begin{tabular}{|c|c|c|c|}
\hline
$\eta$& $E_{\rm bulk}$ [$E_{FG}$]& $E_0$ [$E_{FG}$]& 
$\lambda_s$ [$E_{FG} k_F^{-1}$]\\ \hline \hline
0.00 & -2.3740(3) 			& -2.32(3)  & \phantom{1}6(2) \\ \hline
0.25 & -1.3316(3) 			& -1.31(3)  & \phantom{1}8(2) \\ \hline
0.50 & -0.6766(2) 			& -0.65(2)  & \phantom{1}8(1) \\ \hline
0.75 & -0.2562(2) 			& -0.25(2)  & 11(1)\\ \hline
1.00 &  \phantom{-}0.0233(2)	& \phantom{-}0.03(1) & 11(1) \\ \hline
1.25 &  \phantom{-}0.2149(2)	& \phantom{-}0.22(2) & 12(1) \\ \hline
1.50 &  \phantom{-}0.3523(2)	& \phantom{-}0.34(1) & 13(1)  \\ \hline
\end{tabular}
\label{tab:bulk}
\end{table}

Previous DMC simulations of 2D Fermi gases found that $N=26$ is well 
suited to 
simulate bulk properties of systems in the region studied here 
\cite{gal16}. However, the disk geometry presents more intricate
size-dependent effects.
We investigated how the ground-state energy depends 
on the disk radius ${\cal R}$. In the thermodynamic-limit,
${\cal R}\to \infty$, 
the energy per particle should go to the bulk value.
Since our system 
has hard walls, the energy has a dependence on the ``surface" of the 
disk. Including this surface term, the energy per particle can be fit to
\begin{equation}
\label{eq:eofr}
E_{\rm disk}({\cal R})=E_0 + \frac{\lambda_s}{2\pi{\cal R}},
\end{equation}
where $E_0$ and $\lambda_s$ are constants related to the bulk and 
surface terms, and $\lambda_s/(2\pi {\cal R})$ can be viewed as a surface
tension.

A few words about Eq.~(\ref{eq:eofr}) are in order.
The relation between the thermodynamic properties of
a confined fluid and the shape of the container where it is
confined has been an active field of study.
Our choice was inspired by functional forms (see for example 
Ref.~\cite{kon04}) where, aside from the constant term, thermodynamical 
properties are expressed as functions of the various curvatures of the 
container.
The next correction to this functional form of the energy per particle would 
include a term proportional to ${\cal R}^{-2}$. We found that the inclusion 
of such a term does not significantly improve our description of the ground-
state energy.

In order to determine the number of particles necessary to simulate 
systems in the disk geometry, with controllable size effects, we 
performed simulations with $26 \leqslant N \leqslant 70$, and all 
particles paired, i.e., only even values of $N$. The dependence of 
$E_0$ with the system size was investigated by fitting our data using 
Eq.~(\ref{eq:eofr}) for different intervals of ${\cal R}$ or, 
equivalently, different intervals of $N$. 

\begin{figure}[!htb]
  \centering
  \includegraphics[angle=-90,width=\linewidth]{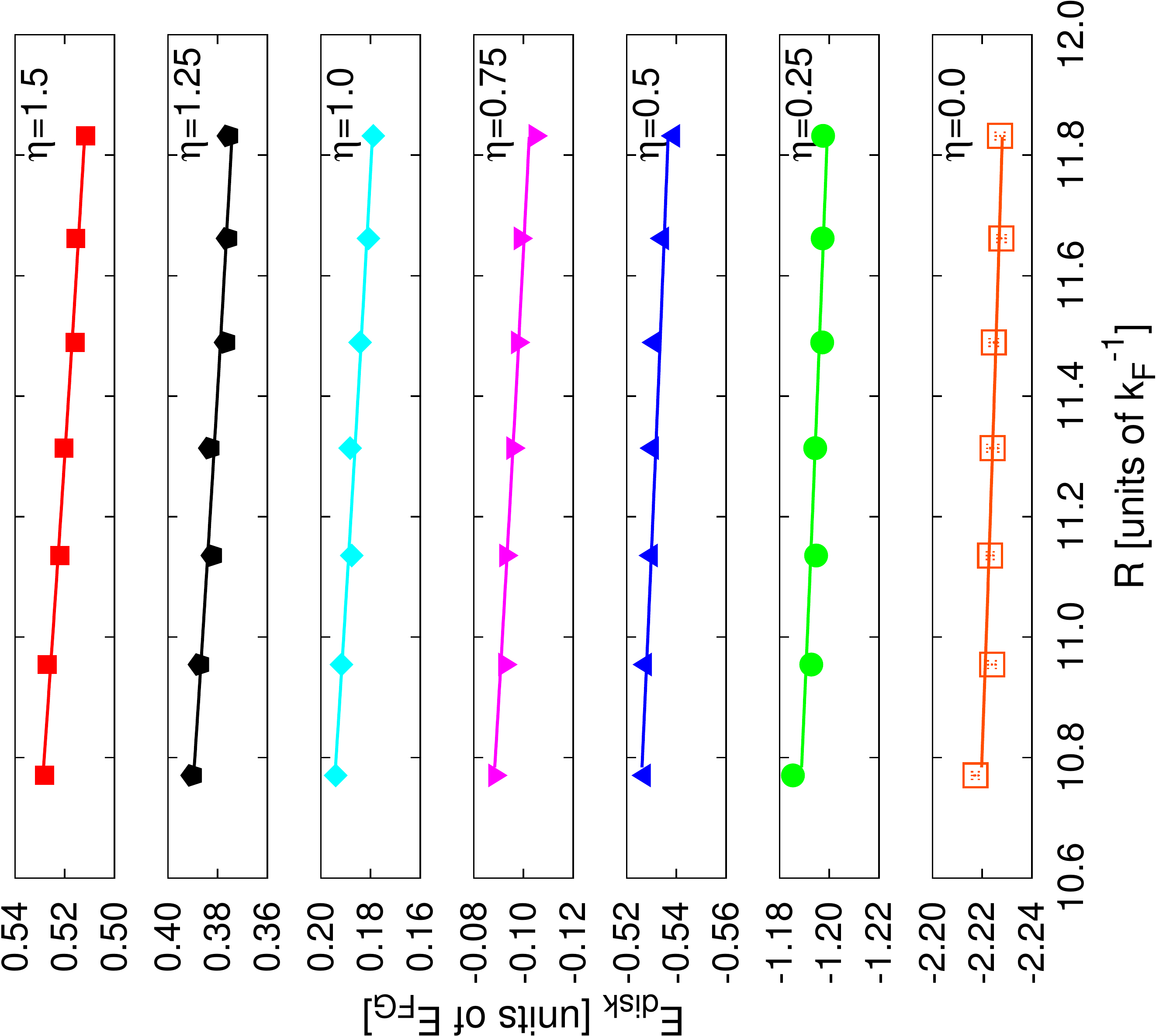}
  \caption{(Color online) Ground-state energy per particle $E_{\rm 
disk}$ as a function of the disk radius ${\cal R}$ for several 
interaction strengths.
The curves correspond to the assumed functional form of 
Eq.~(\ref{eq:eofr}), with the parameters given in Table~\ref{tab:bulk}.
Error 
bars are smaller than the symbols.
}
  \label{fig:eofr}
\end{figure}

We found that fitting the data for $58 \leqslant N \leqslant 70$ 
resulted in a good agreement between $E_{\rm bulk}$ and $E_0$, that is, 
we were able to separate the bulk portion of the energy from the hard 
wall contribution in the disk geometry. The resulting parameters of the 
fitting procedure are summarized in Table~\ref{tab:bulk}, and 
Fig.~\ref{fig:eofr} shows the energy per particle as a function of 
$\cal R$ for all interaction strengths studied in this work.

The $E_0$ values agree with the bulk energies within the error bars, 
except for $\eta=0$ and $\eta=0.5$ (however the differences between the 
values are less than 2\% and 4\%, respectively). As it can be seen in 
Table~\ref{tab:bulk}, the typical uncertainty in $E_0$ is of order 0.01 
$E_F$, independent of the interaction strength. Thus the relative error 
can be quite large for systems where the absolute value of the bulk 
energy is small, as it is observed for $\eta=1.0$. This is an 
improvement if compared to a similar DMC calculation in 3D \cite{mad16} 
which used the same procedure to calculate the ground-state energy per 
particle of a unitary Fermi gas, where the discrepancy between the 
result 
and the known bulk value was $\approx 30$\%.

We point out that this method is not intended to be a precise 
calculation of the bulk energy of these systems. Instead, it is a way 
for us to determine the minimum number of particles needed to simulate 
systems in the disk geometry with controllable size effects. If we had 
naively assumed that the same number of particles used in bulk 
calculations would suffice, $N=26$, then we simply could not rely on 
the results. In our simulations with $26 \leqslant N \leqslant 38$ the 
discrepancies between $E_0$ and $E_{\rm bulk}$ were as large as 50\%, 
and in some cases the uncertainty in $\lambda_s$ was 
bigger than the value itself.
Results with $58 \leqslant N \leqslant 70$ are much more 
well-behaved, and they are within computational capabilities.

It is also noteworthy to mention that the energy contribution of the 
surface term, due to the presence of hard walls, is more significant for 
the BCS side than in the BEC limit (see the $\lambda_s$ values in 
Table~\ref{tab:bulk}). This is expected, since the largest energy 
contribution in the BEC side should be from the binding energy of the 
pairs, Eq.~(\ref{eq:eb}), and they are smaller than the BCS pairs so 
that surface effects are smaller.
One of our goals is to obtain the vortex excitation 
energy, which is the difference between the vortex and the ground-state 
energies. Since both systems have hard walls, we expect that the 
surface effects will tend to cancel.

\subsection{Vortex excitation energy}
\label{sec:vortex}

The energy per particle of the vortex system is obtained using the 
pairing functions of Eq.~(\ref{eq:vortex}). The vortex excitation 
energy is given by the difference between the energy of the vortex and 
ground-state systems, for the same number of particles. We performed 
simulations with $58 \leqslant N \leqslant 70$ and averaged the 
results.

\begin{figure}[!htb]
  \centering
  \includegraphics[angle=-90,width=\linewidth]{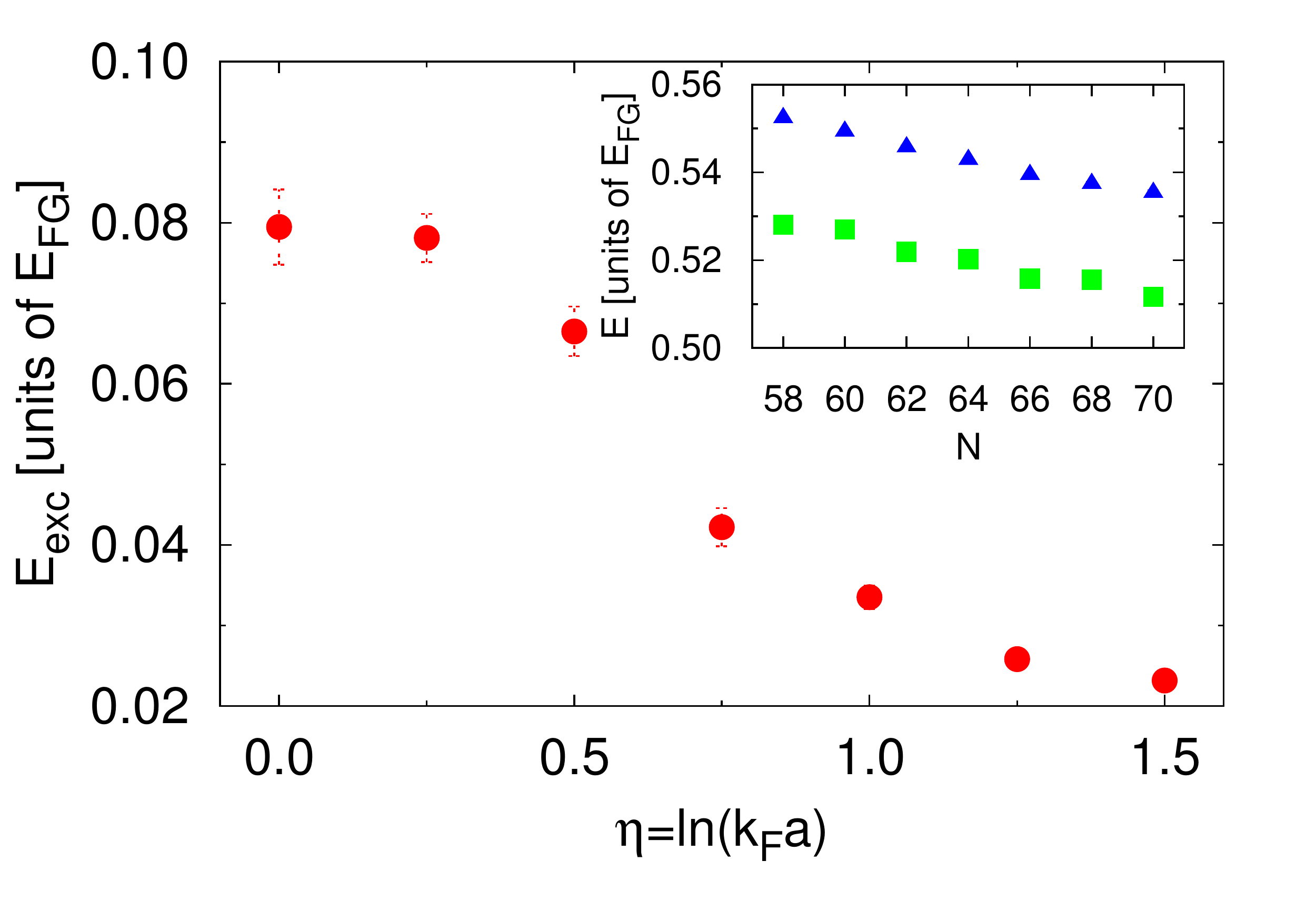}
  \caption{(Color online) Vortex excitation energy per particle $E_{\rm 
exc}$ as a function of the interaction strength $\eta$. The inset 
shows the ground-state (squares) and vortex (triangles) energies per 
particle as a function of the number of particles $N$ for $\eta=1.5$.}
  \label{fig:exc}
\end{figure}

In Fig.~\ref{fig:exc} we show the vortex excitation energy per 
particle as a function of the interaction strength. The energy 
necessary to excite the system to a vortex state increases as we move 
from the BCS to the BEC limit. The inset shows the vortex and
ground-state energies per particle for $\eta=1.5$, although the other 
interaction strengths display the same qualitative behavior.

\subsection{Crossover region}
\label{sec:chempot}

\begin{figure}[!htb]
  \centering
  \includegraphics[angle=-90,width=\linewidth]{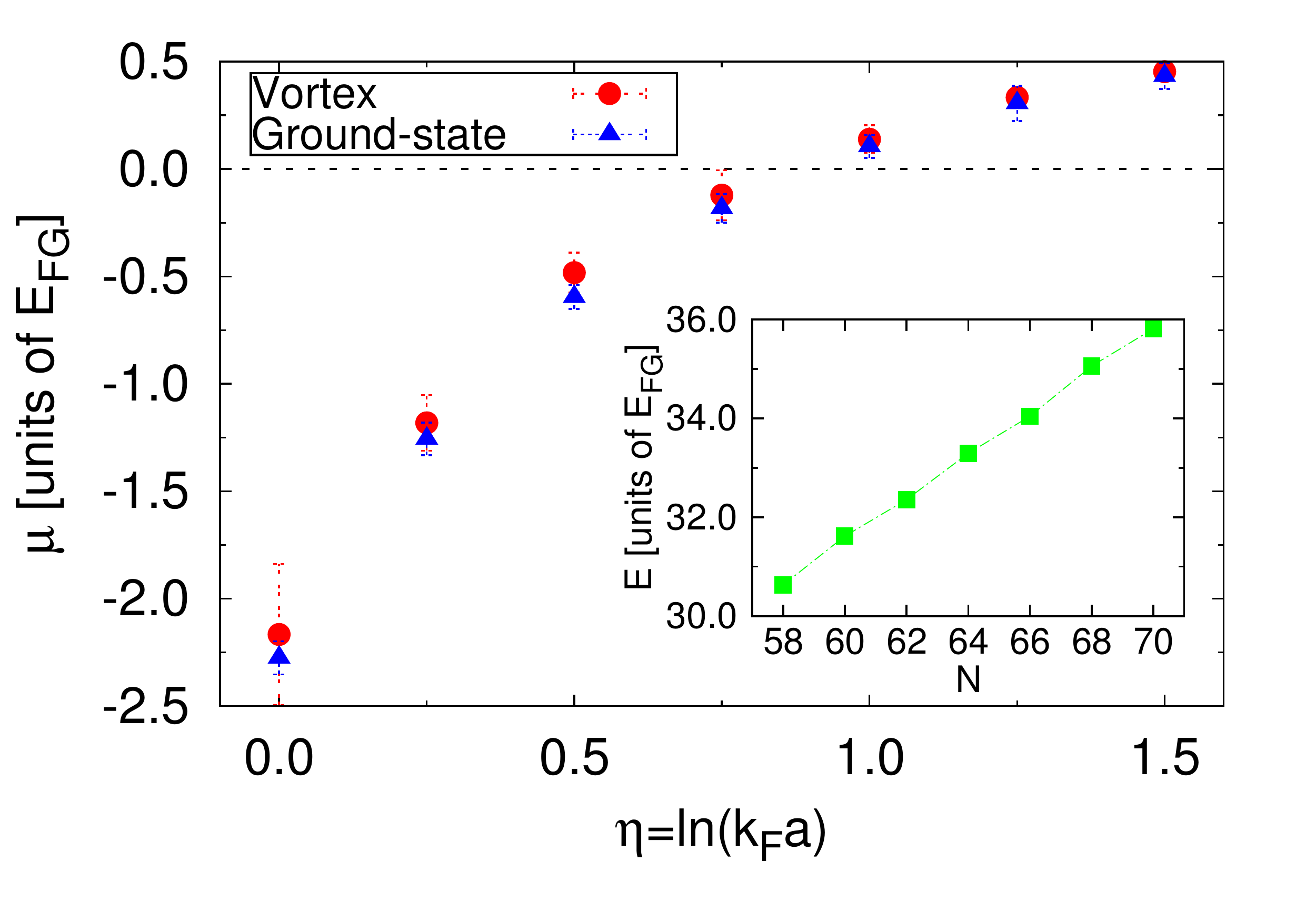}
  \caption{(Color online) Chemical potential of the ground-state 
(triangles) and vortex (circles) as a function of the interaction 
strength. The chemical potential changes sign at $\eta \approx 0.90$ 
for the ground-state, and $\eta \approx 0.85$ for the vortex state.
In the inset we show the total energy as a function of the number of 
particles for the ground-state of $\eta=$ 1.5. Other interaction 
strengths with positive (negative) $\mu$ have positive (negative) 
slopes.}
  \label{fig:mu}
\end{figure}

In 2D, the BCS limit corresponds to $k_F a\gg 1$ and the BEC limit to 
$k_F a\ll 1$, however unlike 3D where the unitarity is signaled by the 
addition of a two-body bound state, there is no equivalent effect with 
two-body sector in 2D.
Nevertheless, we can 
determine the interaction strength for which we can add a pair to the 
system with zero energy cost. The chemical potential $\mu$ can be 
estimated as
\begin{equation}
\label{eq:mu}
\mu = \frac{\partial E}{\partial N} \Bigg|_{{\rm Even\ } N},
\end{equation}
for each interaction strength, where the even number condition implies 
that all particles are paired. For each value of $\eta$ we used a finite 
difference formula to evaluate Eq.~(\ref{eq:mu}), for $58 \leqslant N 
\leqslant 70$ (see Fig.~\ref{fig:mu}).

We found that $\mu=0$ at $\eta \approx 0.90$ for the ground-state of 
the disk. Previous DMC simulations of 2D bulk systems \cite{gal16} 
found 
that the chemical potential changes sign at $\eta \approx 0.65$. 
Although the results differ, most probably due to the different 
geometry employed in this work, it is safe to assume that the 
interaction strength interval $0\leqslant \eta \leqslant 1.5$ 
encompasses the BEC-BCS crossover region. The chemical potential of the 
vortex state is higher than the ground-state, as expected, thus $\mu=0$ is
at a smaller interaction strength, $\eta \approx 0.85$.

\subsection{Density profile}
\label{sec:density}

We calculated the density profile ${\cal D}(\rho)$ along the radial 
direction $\rho$ for both the vortex and ground-state systems. The 
normalization is such that
\begin{equation}
\int {\cal D}(\rho) d^2r=1,
\end{equation}
where the integral is performed over the area of the disk.
The results are obtained using the extrapolation procedure of 
Eq.~(\ref{eq:ext}), which combines both VMC and DMC runs. It is 
noteworthy to point out that, although the densities observed in VMC 
and DMC simulations differ, they are much closer than previous results 
in 3D \cite{mad16}. In that calculation it was needed to explicitly 
include a one-body term in the wave function to maximize the density 
overlap between DMC and VMC runs, whereas in this work no such term was 
employed. 

\begin{figure}[!htb]
  \centering
  \includegraphics[angle=-90,width=\linewidth]{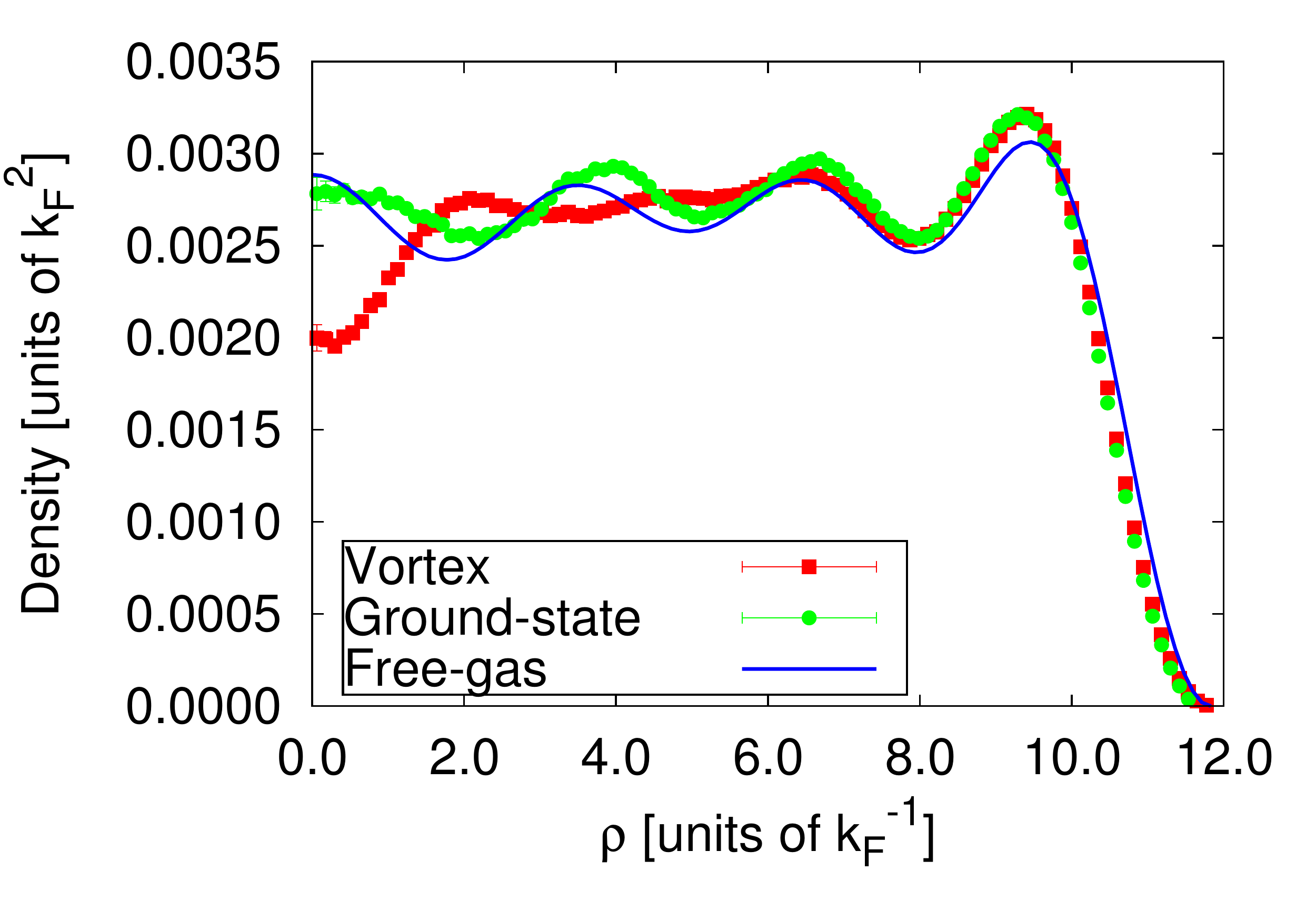}
  \caption{(Color online) Density profile along the radial direction 
$\rho$ of the vortex (red squares) and ground-state (green circles)
for $N=70$ 
and $\eta=1.5$. Although there is a density suppression at the vortex 
core of $\approx$ 30\%, the density is non-zero at the center of the 
disk.
We also plot the analytical result for the ground-state density of the free-
gas in a disk (blue curve), which presents oscillations due to the presence 
of hard-walls.
}
  \label{fig:dens15}
\end{figure}

Figure \ref{fig:dens15} shows the density profile of both the vortex and 
ground-state systems for $N=70$ and $\eta=1.5$. The oscillations in the 
density profiles are much more pronounced than in a similar DMC 
calculation of a unitary Fermi gas in 3D \cite{mad16}.
In this 3D calculation a cylindrical geometry was employed, with hard 
walls and periodic boundary conditions along the axis of the cylinder.
The density profiles were obtained by averaging the results over the 
$z$ direction of the axis of the cylinder, 
we therefore expect more fluctuations in 2D where 
the particles are 
confined to a plane.
For the ground-state, the density oscillations are surface effects. They are 
present in both the interacting and non-interacting systems, as it can be 
seen in Fig.~\ref{fig:dens15}.

\begin{figure}[!htb]
  \centering
  \includegraphics[angle=-90,width=\linewidth]{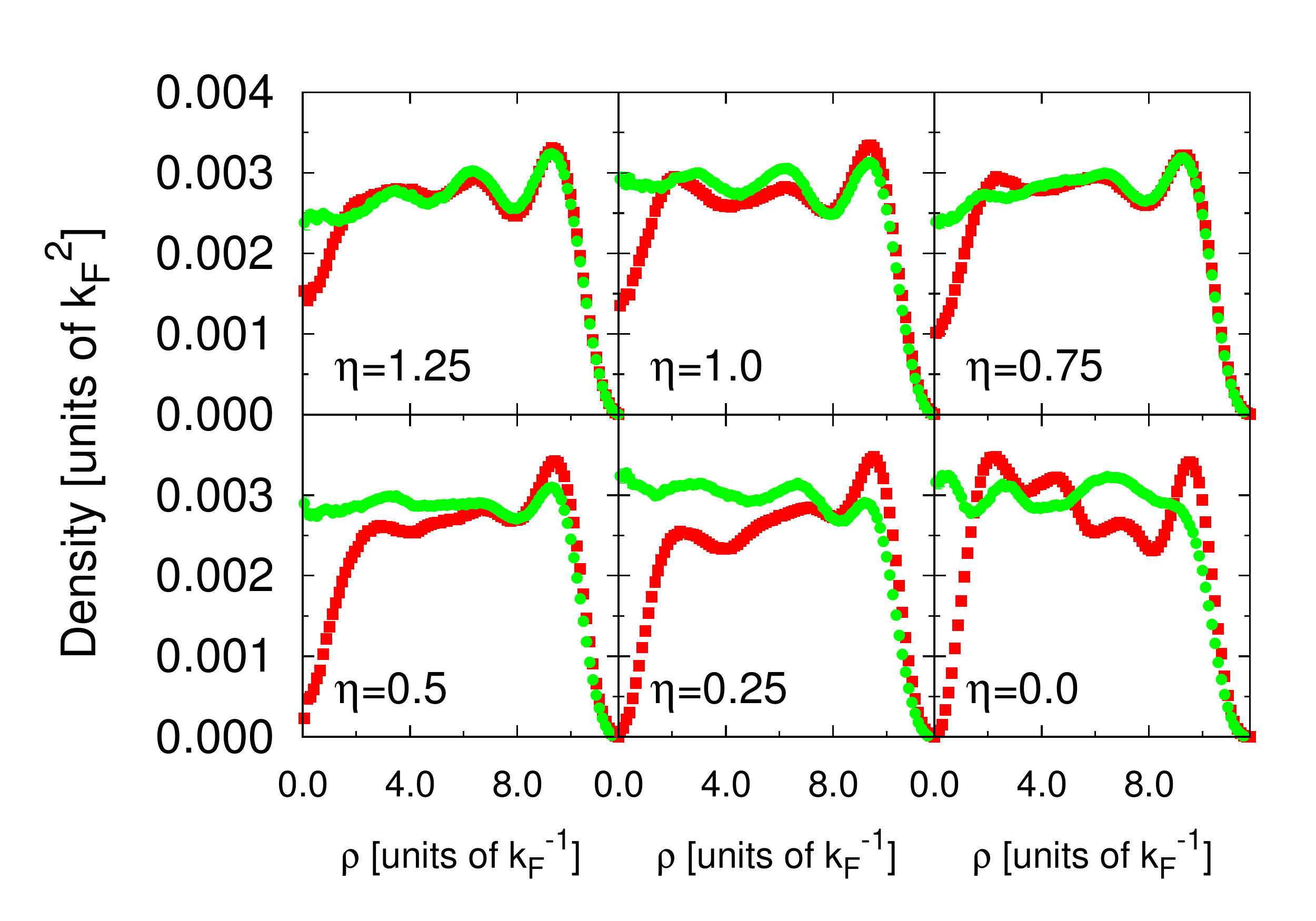}
  \caption{(Color online) Density profile along the radial direction 
$\rho$ of the vortex (red squares) and ground-state (green circles)
for $N=70$ 
and $0\leqslant \eta \leqslant 1.25$. It is interesting to observe that 
the density at the vortex core diminishes as we go from the BCS to the 
BEC limit, and at $\eta \leqslant 0.25$ the core is completely 
depleted.
}
  \label{fig:dens}
\end{figure}

In Fig.~\ref{fig:dens} we show the density profiles of the other 
interaction strengths studied in this work, $0\leqslant \eta \leqslant 
1.25$. We found that the density depletion at the vortex core goes from 
$\approx$ 30\% at $\eta=1.5$ to a completely depleted core at $\eta 
\leqslant$ 0.25.

The regions close to the walls exhibit a characteristic behavior due to 
the hard wall condition we imposed, as it can be seen in 
Figs.~\ref{fig:dens15} and \ref{fig:dens}.
In order to estimate the number of particles outside this region, we can 
define the particle number a distance $R$ from the center of the disk as
\begin{eqnarray}
\mathcal{N}(R)=N\int_0^{2\pi}d\varphi \int_0^R d\rho \ \rho \ 
\mathcal{D}(\rho).
\end{eqnarray}
For the case of Figs.~\ref{fig:dens15} and \ref{fig:dens} where 
$N=70$, if we 
set $R\sim 8 \ k_F^{-1}$, $\mathcal{N}$ is approximately between 40 and 
45 for the ground-state, and between 35 and 40 for the vortex systems. 
Hence the number of particles in this regime is larger than the usual 
value of $N=26$ employed in bulk systems \cite{gal16}.

Additionally, we performed simulations of the vortex systems with an odd 
number of particles, i.e., one unpaired particle was added to a fully 
paired system, Eq.~(\ref{eq:unpaired}) with $q=34$, $u=1$, and $d=0$.
We set its angular momentum to zero, Eq.~(\ref{eq:spdisk}) with $\nu=0$ 
and $p=1$.
In the BEC limit we observed a non-vanishing density at the center of 
the disk, which suggests that the unpaired particle fills the empty 
vortex core region. On the other hand, in the BCS limit the density 
close to the wall increased, while the density at the origin was 
unchanged.
We chose a qualitative discussion of this phenomenon because the 
required variance for a detailed optimization is beyond the scope of 
this work.
Future calculations should include quantities such as the
one-body density matrix, which may contribute to an accurate 
quantitative approach.

\subsection{Vortex core size}
\label{sec:current}

The probability current density operator can be written as
\begin{eqnarray}
\label{eq:jop}
\textbf{J}(\textbf{r})=\frac{1}{2N}\sum_{j=1}^{N}
\left[
\textbf{v}_j \delta^2(\textbf{r}-\textbf{r}_j)+
\delta^2(\textbf{r}-\textbf{r}_j)\textbf{v}_j
\right], 
\end{eqnarray} 
where the velocity operator is 
$\textbf{v}_j=\textbf{p}_j/m\to -i\hbar\nabla_j/m$.
We are interested in the angular component as a function of the radial 
coordinate, $J_\varphi (\rho)$, because the position of its maximum can 
be used as an estimate of the vortex core size, $J_{\rm max} \equiv 
J_\varphi(\rho=\xi)$.

\begin{figure}[!htb]
  \centering
  \includegraphics[angle=-90,width=\linewidth]{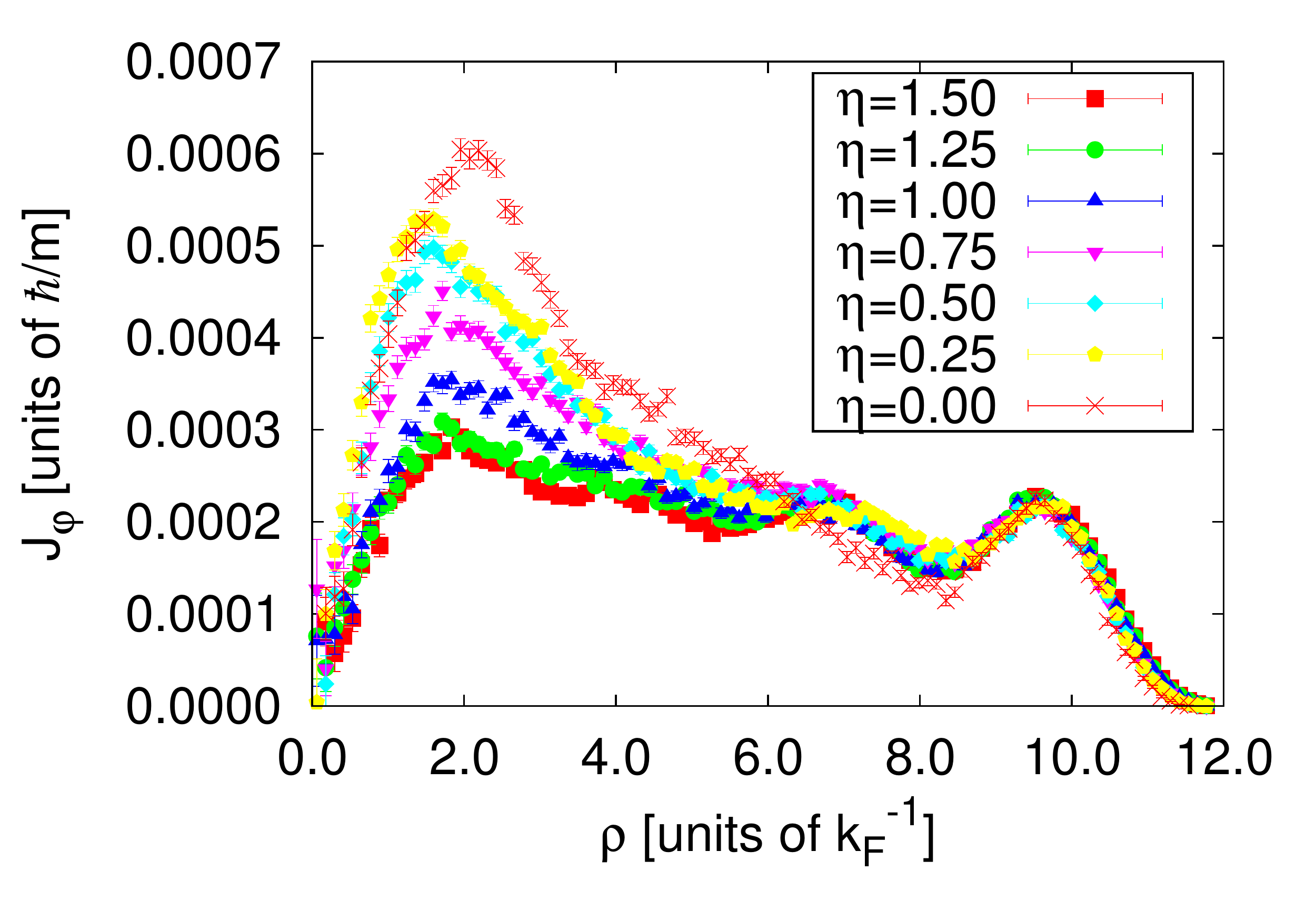}
  \caption{(Color online) Angular component of the probability current 
$J_\varphi$ as a function of the radial coordinate $\rho$ for several 
interaction strengths $\eta$. The position of its maximum provides an 
estimate of the vortex core size.}
  \label{fig:curr}
\end{figure}

We followed the extrapolation procedure of Eq.~(\ref{eq:ext}). 
Figure~\ref{fig:curr} shows $J_\varphi (\rho)$ for $N=70$ and $0 
\leqslant \eta \leqslant 1.5$.
The maximum of the current increases as we go from the BCS to the BEC 
limit, its value at the BEC side, $\eta=0$, being more than twice 
$J_{\rm 
max}$ at the BCS side, $\eta=1.5$.
The position of the maximum is between $\xi=$ 1.7 and 1.8 $k_F^{-1}$ at 
the BCS side of the crossover, i.e., $0.75 \leqslant \eta \leqslant 
1.5$; at the BEC side, $\eta=0.25$ and 0.5, $\xi \sim 1.6$ $k_F^{-1}$. 
The case $\eta=0$ moves away from the trend of a smaller core as we go 
from the BCS to the BEC limit, with $\xi=2.0$ $k_F^{-1}$.
It is unclear if $\xi$ or $J_{\rm max}$ depend on the disk radius 
${\cal R}$, because the ${\cal R}$ values are closely spaced for $58 
\leqslant N \leqslant 70$, and no significant difference was observed 
in the maximum as we varied $N$. Nevertheless, the relative results 
contribute to understanding how the vortex core evolves over the
BEC-BCS crossover.

The wave function that we employed for the vortex state is an eigenstate of 
the total angular momentum operator.
Since this operator commutes with the Hamiltonian, the diffusion procedure 
does not change the eigenvalue of the state.
In addition, the calculation of the probability current density operator 
allowed us to verify that the vortex corresponds to a $N\hbar/2$ total 
angular momentum state in a straightforward way. The angular momentum can be 
written as
\begin{eqnarray}
\textbf{L}=m \int (\textbf{r} \times \textbf{J}) d^2\textbf{r}, 
\end{eqnarray}
and the component of interest is
\begin{eqnarray}
\label{eq:Lz}
L_z=2\pi m \int \rho^2 J_\varphi(\rho) d\rho .
\end{eqnarray}
In our definition of the probability current density operator, we divide by 
the number of particles $N$, see Eq.~(\ref{eq:jop}). Thus, the evaluation of 
$L_z$ using Eq.~(\ref{eq:Lz}) should yield $\hbar/2$. We verified that,
for all interaction strengths,
this is in
agreement with our 
simulations.

\section{Summary}
\label{sec:sum}

We have investigated several properties of vortices in 2D Fermi gases 
over the BEC-BCS crossover region. We dedicated a considerable portion 
of this work to carefully understand and control size effects in the 
disk geometry, since it is very convenient to simulating a single 
vortex.
Given that we were interested in the evolution of the properties in 
the BEC-BCS crossover, determining the crossover region was important 
to verify that the interaction strengths studied in this work span 
the crossover.

The vortex excitation energies and the density profiles are
quantities that can be compared with 
experiments, once they become available. Interestingly, the observed 
density depletion of the vortex core goes from $\approx$ 30\% at the 
BCS side, $\eta=1.5$, to an empty core for $\eta \leqslant 0.25$, at 
the BEC limit. In 3D, Bogoliubov-de Gennes theory has been used to 
calculate the density suppression at the vortex core throughout the 
BEC-BCS crossover \cite{bul03,sen06,sim13}. Similar calculations in 2D could 
be compared to our findings
\footnote{After our calculations were done, it was pointed out to us that 
pseudogap phenomena occurring in 2D and 3D Fermi gases can be related in a 
universal way through a variable that spans the BEC-BCS crossover 
\cite{mar15}. Further studies are necessary to determine if this universality 
holds for other quantities, such as the density and the probability current 
density per particle. This would provide a very clean way of comparing 2D and 
3D results.}.
Also, determining the probability 
current was essential to investigate the changes in the vortex core 
throughout the crossover region.

In 3D the interplay between experiments, theory, and simulations led to 
rapid advances in our comprehension of cold Fermi gases. Hopefully, our 
results will motivate experiments to increase our understanding of 
vortices in 2D Fermi gases.

\begin{acknowledgments}
We would like to thank G. C. Strinati for useful discussions.
This work was supported by the National Science
Foundation under grant PHY-1404405. This work used the Extreme Science and 
Engineering Discovery Environment 
(XSEDE), which is supported by National Science Foundation grant number 
ACI-1053575.
The work of S.G. was supported by the NUCLEI SciDAC program, the U.S. DOE 
under 
Contract No. DE-AC52-06NA25396, and the LANL LDRD program. We also used 
resources provided by NERSC, which is supported by the U.S. DOE under 
Contract 
No. DE-AC02-05CH11231.
\end{acknowledgments}
%\appendix
%\label{sec:appendix}
\bibliography{article}% Produces the bibliography via BibTeX.

\begin{thebibliography}{47}%
\makeatletter
\providecommand \@ifxundefined [1]{%
 \@ifx{#1\undefined}
}%
\providecommand \@ifnum [1]{%
 \ifnum #1\expandafter \@firstoftwo
 \else \expandafter \@secondoftwo
 \fi
}%
\providecommand \@ifx [1]{%
 \ifx #1\expandafter \@firstoftwo
 \else \expandafter \@secondoftwo
 \fi
}%
\providecommand \natexlab [1]{#1}%
\providecommand \enquote  [1]{``#1''}%
\providecommand \bibnamefont  [1]{#1}%
\providecommand \bibfnamefont [1]{#1}%
\providecommand \citenamefont [1]{#1}%
\providecommand \href@noop [0]{\@secondoftwo}%
\providecommand \href [0]{\begingroup \@sanitize@url \@href}%
\providecommand \@href[1]{\@@startlink{#1}\@@href}%
\providecommand \@@href[1]{\endgroup#1\@@endlink}%
\providecommand \@sanitize@url [0]{\catcode `\\12\catcode `\$12\catcode
  `\&12\catcode `\#12\catcode `\^12\catcode `\_12\catcode `\%12\relax}%
\providecommand \@@startlink[1]{}%
\providecommand \@@endlink[0]{}%
\providecommand \url  [0]{\begingroup\@sanitize@url \@url }%
\providecommand \@url [1]{\endgroup\@href {#1}{\urlprefix }}%
\providecommand \urlprefix  [0]{URL }%
\providecommand \Eprint [0]{\href }%
\providecommand \doibase [0]{http://dx.doi.org/}%
\providecommand \selectlanguage [0]{\@gobble}%
\providecommand \bibinfo  [0]{\@secondoftwo}%
\providecommand \bibfield  [0]{\@secondoftwo}%
\providecommand \translation [1]{[#1]}%
\providecommand \BibitemOpen [0]{}%
\providecommand \bibitemStop [0]{}%
\providecommand \bibitemNoStop [0]{.\EOS\space}%
\providecommand \EOS [0]{\spacefactor3000\relax}%
\providecommand \BibitemShut  [1]{\csname bibitem#1\endcsname}%
\let\auto@bib@innerbib\@empty
%</preamble>
\bibitem [{\citenamefont {Giorgini}\ \emph {et~al.}(2008)\citenamefont
  {Giorgini}, \citenamefont {Pitaevskii},\ and\ \citenamefont
  {Stringari}}]{gio08}%
  \BibitemOpen
  \bibfield  {author} {\bibinfo {author} {\bibfnamefont {S.}~\bibnamefont
  {Giorgini}}, \bibinfo {author} {\bibfnamefont {L.~P.}\ \bibnamefont
  {Pitaevskii}}, \ and\ \bibinfo {author} {\bibfnamefont {S.}~\bibnamefont
  {Stringari}},\ }\href {\doibase 10.1103/RevModPhys.80.1215} {\bibfield
  {journal} {\bibinfo  {journal} {Rev. Mod. Phys.}\ }\textbf {\bibinfo {volume}
  {80}},\ \bibinfo {pages} {1215} (\bibinfo {year} {2008})}\BibitemShut
  {NoStop}%
\bibitem [{\citenamefont {Bloch}\ \emph {et~al.}(2008)\citenamefont {Bloch},
  \citenamefont {Dalibard},\ and\ \citenamefont {Zwerger}}]{blo08}%
  \BibitemOpen
  \bibfield  {author} {\bibinfo {author} {\bibfnamefont {I.}~\bibnamefont
  {Bloch}}, \bibinfo {author} {\bibfnamefont {J.}~\bibnamefont {Dalibard}}, \
  and\ \bibinfo {author} {\bibfnamefont {W.}~\bibnamefont {Zwerger}},\ }\href
  {\doibase 10.1103/RevModPhys.80.885} {\bibfield  {journal} {\bibinfo
  {journal} {Rev. Mod. Phys.}\ }\textbf {\bibinfo {volume} {80}},\ \bibinfo
  {pages} {885} (\bibinfo {year} {2008})}\BibitemShut {NoStop}%
\bibitem [{\citenamefont {Randeria}\ \emph {et~al.}(1989)\citenamefont
  {Randeria}, \citenamefont {Duan},\ and\ \citenamefont {Shieh}}]{ran89}%
  \BibitemOpen
  \bibfield  {author} {\bibinfo {author} {\bibfnamefont {M.}~\bibnamefont
  {Randeria}}, \bibinfo {author} {\bibfnamefont {J.-M.}\ \bibnamefont {Duan}},
  \ and\ \bibinfo {author} {\bibfnamefont {L.-Y.}\ \bibnamefont {Shieh}},\
  }\href {\doibase 10.1103/PhysRevLett.62.981} {\bibfield  {journal} {\bibinfo
  {journal} {Phys. Rev. Lett.}\ }\textbf {\bibinfo {volume} {62}},\ \bibinfo
  {pages} {981} (\bibinfo {year} {1989})}\BibitemShut {NoStop}%
\bibitem [{\citenamefont {Randeria}\ \emph {et~al.}(1990)\citenamefont
  {Randeria}, \citenamefont {Duan},\ and\ \citenamefont {Shieh}}]{ran90}%
  \BibitemOpen
  \bibfield  {author} {\bibinfo {author} {\bibfnamefont {M.}~\bibnamefont
  {Randeria}}, \bibinfo {author} {\bibfnamefont {J.-M.}\ \bibnamefont {Duan}},
  \ and\ \bibinfo {author} {\bibfnamefont {L.-Y.}\ \bibnamefont {Shieh}},\
  }\href {\doibase 10.1103/PhysRevB.41.327} {\bibfield  {journal} {\bibinfo
  {journal} {Phys. Rev. B}\ }\textbf {\bibinfo {volume} {41}},\ \bibinfo
  {pages} {327} (\bibinfo {year} {1990})}\BibitemShut {NoStop}%
\bibitem [{\citenamefont {Petrov}\ \emph {et~al.}(2003)\citenamefont {Petrov},
  \citenamefont {Baranov},\ and\ \citenamefont {Shlyapnikov}}]{pet03}%
  \BibitemOpen
  \bibfield  {author} {\bibinfo {author} {\bibfnamefont {D.~S.}\ \bibnamefont
  {Petrov}}, \bibinfo {author} {\bibfnamefont {M.~A.}\ \bibnamefont {Baranov}},
  \ and\ \bibinfo {author} {\bibfnamefont {G.~V.}\ \bibnamefont
  {Shlyapnikov}},\ }\href {\doibase 10.1103/PhysRevA.67.031601} {\bibfield
  {journal} {\bibinfo  {journal} {Phys. Rev. A}\ }\textbf {\bibinfo {volume}
  {67}},\ \bibinfo {pages} {031601} (\bibinfo {year} {2003})}\BibitemShut
  {NoStop}%
\bibitem [{\citenamefont {Martikainen}\ and\ \citenamefont
  {T\"orm\"a}(2005)}]{mar05}%
  \BibitemOpen
  \bibfield  {author} {\bibinfo {author} {\bibfnamefont {J.-P.}\ \bibnamefont
  {Martikainen}}\ and\ \bibinfo {author} {\bibfnamefont {P.}~\bibnamefont
  {T\"orm\"a}},\ }\href {\doibase 10.1103/PhysRevLett.95.170407} {\bibfield
  {journal} {\bibinfo  {journal} {Phys. Rev. Lett.}\ }\textbf {\bibinfo
  {volume} {95}},\ \bibinfo {pages} {170407} (\bibinfo {year}
  {2005})}\BibitemShut {NoStop}%
\bibitem [{\citenamefont {Tempere}\ \emph {et~al.}(2007)\citenamefont
  {Tempere}, \citenamefont {Wouters},\ and\ \citenamefont {Devreese}}]{tem07}%
  \BibitemOpen
  \bibfield  {author} {\bibinfo {author} {\bibfnamefont {J.}~\bibnamefont
  {Tempere}}, \bibinfo {author} {\bibfnamefont {M.}~\bibnamefont {Wouters}}, \
  and\ \bibinfo {author} {\bibfnamefont {J.~T.}\ \bibnamefont {Devreese}},\
  }\href {\doibase 10.1103/PhysRevB.75.184526} {\bibfield  {journal} {\bibinfo
  {journal} {Phys. Rev. B}\ }\textbf {\bibinfo {volume} {75}},\ \bibinfo
  {pages} {184526} (\bibinfo {year} {2007})}\BibitemShut {NoStop}%
\bibitem [{\citenamefont {Zhang}\ \emph {et~al.}(2008)\citenamefont {Zhang},
  \citenamefont {Lin},\ and\ \citenamefont {Duan}}]{zha08}%
  \BibitemOpen
  \bibfield  {author} {\bibinfo {author} {\bibfnamefont {W.}~\bibnamefont
  {Zhang}}, \bibinfo {author} {\bibfnamefont {G.-D.}\ \bibnamefont {Lin}}, \
  and\ \bibinfo {author} {\bibfnamefont {L.-M.}\ \bibnamefont {Duan}},\ }\href
  {\doibase 10.1103/PhysRevA.78.043617} {\bibfield  {journal} {\bibinfo
  {journal} {Phys. Rev. A}\ }\textbf {\bibinfo {volume} {78}},\ \bibinfo
  {pages} {043617} (\bibinfo {year} {2008})}\BibitemShut {NoStop}%
\bibitem [{\citenamefont {Martiyanov}\ \emph {et~al.}(2010)\citenamefont
  {Martiyanov}, \citenamefont {Makhalov},\ and\ \citenamefont
  {Turlapov}}]{kir10}%
  \BibitemOpen
  \bibfield  {author} {\bibinfo {author} {\bibfnamefont {K.}~\bibnamefont
  {Martiyanov}}, \bibinfo {author} {\bibfnamefont {V.}~\bibnamefont
  {Makhalov}}, \ and\ \bibinfo {author} {\bibfnamefont {A.}~\bibnamefont
  {Turlapov}},\ }\href {\doibase 10.1103/PhysRevLett.105.030404} {\bibfield
  {journal} {\bibinfo  {journal} {Phys. Rev. Lett.}\ }\textbf {\bibinfo
  {volume} {105}},\ \bibinfo {pages} {030404} (\bibinfo {year}
  {2010})}\BibitemShut {NoStop}%
\bibitem [{\citenamefont {Orel}\ \emph {et~al.}(2011)\citenamefont {Orel},
  \citenamefont {Dyke}, \citenamefont {Delehaye}, \citenamefont {Vale},\ and\
  \citenamefont {Hu}}]{ore11}%
  \BibitemOpen
  \bibfield  {author} {\bibinfo {author} {\bibfnamefont {A.~A.}\ \bibnamefont
  {Orel}}, \bibinfo {author} {\bibfnamefont {P.}~\bibnamefont {Dyke}}, \bibinfo
  {author} {\bibfnamefont {M.}~\bibnamefont {Delehaye}}, \bibinfo {author}
  {\bibfnamefont {C.~J.}\ \bibnamefont {Vale}}, \ and\ \bibinfo {author}
  {\bibfnamefont {H.}~\bibnamefont {Hu}},\ }\href
  {http://stacks.iop.org/1367-2630/13/i=11/a=113032} {\bibfield  {journal}
  {\bibinfo  {journal} {New Journal of Physics}\ }\textbf {\bibinfo {volume}
  {13}},\ \bibinfo {pages} {113032} (\bibinfo {year} {2011})}\BibitemShut
  {NoStop}%
\bibitem [{\citenamefont {Makhalov}\ \emph {et~al.}(2014)\citenamefont
  {Makhalov}, \citenamefont {Martiyanov},\ and\ \citenamefont
  {Turlapov}}]{mak14}%
  \BibitemOpen
  \bibfield  {author} {\bibinfo {author} {\bibfnamefont {V.}~\bibnamefont
  {Makhalov}}, \bibinfo {author} {\bibfnamefont {K.}~\bibnamefont
  {Martiyanov}}, \ and\ \bibinfo {author} {\bibfnamefont {A.}~\bibnamefont
  {Turlapov}},\ }\href {\doibase 10.1103/PhysRevLett.112.045301} {\bibfield
  {journal} {\bibinfo  {journal} {Phys. Rev. Lett.}\ }\textbf {\bibinfo
  {volume} {112}},\ \bibinfo {pages} {045301} (\bibinfo {year}
  {2014})}\BibitemShut {NoStop}%
\bibitem [{\citenamefont {Bertaina}\ and\ \citenamefont
  {Giorgini}(2011)}]{ber11}%
  \BibitemOpen
  \bibfield  {author} {\bibinfo {author} {\bibfnamefont {G.}~\bibnamefont
  {Bertaina}}\ and\ \bibinfo {author} {\bibfnamefont {S.}~\bibnamefont
  {Giorgini}},\ }\href {\doibase 10.1103/PhysRevLett.106.110403} {\bibfield
  {journal} {\bibinfo  {journal} {Phys. Rev. Lett.}\ }\textbf {\bibinfo
  {volume} {106}},\ \bibinfo {pages} {110403} (\bibinfo {year}
  {2011})}\BibitemShut {NoStop}%
\bibitem [{\citenamefont {Galea}\ \emph {et~al.}(2016)\citenamefont {Galea},
  \citenamefont {Dawkins}, \citenamefont {Gandolfi},\ and\ \citenamefont
  {Gezerlis}}]{gal16}%
  \BibitemOpen
  \bibfield  {author} {\bibinfo {author} {\bibfnamefont {A.}~\bibnamefont
  {Galea}}, \bibinfo {author} {\bibfnamefont {H.}~\bibnamefont {Dawkins}},
  \bibinfo {author} {\bibfnamefont {S.}~\bibnamefont {Gandolfi}}, \ and\
  \bibinfo {author} {\bibfnamefont {A.}~\bibnamefont {Gezerlis}},\ }\href
  {\doibase 10.1103/PhysRevA.93.023602} {\bibfield  {journal} {\bibinfo
  {journal} {Phys. Rev. A}\ }\textbf {\bibinfo {volume} {93}},\ \bibinfo
  {pages} {023602} (\bibinfo {year} {2016})}\BibitemShut {NoStop}%
\bibitem [{\citenamefont {Shi}\ \emph {et~al.}(2015)\citenamefont {Shi},
  \citenamefont {Chiesa},\ and\ \citenamefont {Zhang}}]{shi15}%
  \BibitemOpen
  \bibfield  {author} {\bibinfo {author} {\bibfnamefont {H.}~\bibnamefont
  {Shi}}, \bibinfo {author} {\bibfnamefont {S.}~\bibnamefont {Chiesa}}, \ and\
  \bibinfo {author} {\bibfnamefont {S.}~\bibnamefont {Zhang}},\ }\href
  {\doibase 10.1103/PhysRevA.92.033603} {\bibfield  {journal} {\bibinfo
  {journal} {Phys. Rev. A}\ }\textbf {\bibinfo {volume} {92}},\ \bibinfo
  {pages} {033603} (\bibinfo {year} {2015})}\BibitemShut {NoStop}%
\bibitem [{\citenamefont {Anderson}\ and\ \citenamefont {Drut}(2015)}]{and15}%
  \BibitemOpen
  \bibfield  {author} {\bibinfo {author} {\bibfnamefont {E.~R.}\ \bibnamefont
  {Anderson}}\ and\ \bibinfo {author} {\bibfnamefont {J.~E.}\ \bibnamefont
  {Drut}},\ }\href {\doibase 10.1103/PhysRevLett.115.115301} {\bibfield
  {journal} {\bibinfo  {journal} {Phys. Rev. Lett.}\ }\textbf {\bibinfo
  {volume} {115}},\ \bibinfo {pages} {115301} (\bibinfo {year}
  {2015})}\BibitemShut {NoStop}%
\bibitem [{\citenamefont {Rammelm\"uller}\ \emph {et~al.}(2016)\citenamefont
  {Rammelm\"uller}, \citenamefont {Porter},\ and\ \citenamefont
  {Drut}}]{ram16}%
  \BibitemOpen
  \bibfield  {author} {\bibinfo {author} {\bibfnamefont {L.}~\bibnamefont
  {Rammelm\"uller}}, \bibinfo {author} {\bibfnamefont {W.~J.}\ \bibnamefont
  {Porter}}, \ and\ \bibinfo {author} {\bibfnamefont {J.~E.}\ \bibnamefont
  {Drut}},\ }\href {\doibase 10.1103/PhysRevA.93.033639} {\bibfield  {journal}
  {\bibinfo  {journal} {Phys. Rev. A}\ }\textbf {\bibinfo {volume} {93}},\
  \bibinfo {pages} {033639} (\bibinfo {year} {2016})}\BibitemShut {NoStop}%
\bibitem [{\citenamefont {Luo}\ \emph {et~al.}(2016)\citenamefont {Luo},
  \citenamefont {Berger},\ and\ \citenamefont {Drut}}]{luo16}%
  \BibitemOpen
  \bibfield  {author} {\bibinfo {author} {\bibfnamefont {Z.}~\bibnamefont
  {Luo}}, \bibinfo {author} {\bibfnamefont {C.~E.}\ \bibnamefont {Berger}}, \
  and\ \bibinfo {author} {\bibfnamefont {J.~E.}\ \bibnamefont {Drut}},\ }\href
  {\doibase 10.1103/PhysRevA.93.033604} {\bibfield  {journal} {\bibinfo
  {journal} {Phys. Rev. A}\ }\textbf {\bibinfo {volume} {93}},\ \bibinfo
  {pages} {033604} (\bibinfo {year} {2016})}\BibitemShut {NoStop}%
\bibitem [{\citenamefont {Bulgac}\ and\ \citenamefont {Yu}(2003)}]{bul03}%
  \BibitemOpen
  \bibfield  {author} {\bibinfo {author} {\bibfnamefont {A.}~\bibnamefont
  {Bulgac}}\ and\ \bibinfo {author} {\bibfnamefont {Y.}~\bibnamefont {Yu}},\
  }\href {\doibase 10.1103/PhysRevLett.91.190404} {\bibfield  {journal}
  {\bibinfo  {journal} {Phys. Rev. Lett.}\ }\textbf {\bibinfo {volume} {91}},\
  \bibinfo {pages} {190404} (\bibinfo {year} {2003})}\BibitemShut {NoStop}%
\bibitem [{\citenamefont {Sensarma}\ \emph {et~al.}(2006)\citenamefont
  {Sensarma}, \citenamefont {Randeria},\ and\ \citenamefont {Ho}}]{sen06}%
  \BibitemOpen
  \bibfield  {author} {\bibinfo {author} {\bibfnamefont {R.}~\bibnamefont
  {Sensarma}}, \bibinfo {author} {\bibfnamefont {M.}~\bibnamefont {Randeria}},
  \ and\ \bibinfo {author} {\bibfnamefont {T.-L.}\ \bibnamefont {Ho}},\ }\href
  {\doibase 10.1103/PhysRevLett.96.090403} {\bibfield  {journal} {\bibinfo
  {journal} {Phys. Rev. Lett.}\ }\textbf {\bibinfo {volume} {96}},\ \bibinfo
  {pages} {090403} (\bibinfo {year} {2006})}\BibitemShut {NoStop}%
\bibitem [{\citenamefont {Simonucci}\ \emph {et~al.}(2013)\citenamefont
  {Simonucci}, \citenamefont {Pieri},\ and\ \citenamefont {Strinati}}]{sim13}%
  \BibitemOpen
  \bibfield  {author} {\bibinfo {author} {\bibfnamefont {S.}~\bibnamefont
  {Simonucci}}, \bibinfo {author} {\bibfnamefont {P.}~\bibnamefont {Pieri}}, \
  and\ \bibinfo {author} {\bibfnamefont {G.~C.}\ \bibnamefont {Strinati}},\
  }\href {\doibase 10.1103/PhysRevB.87.214507} {\bibfield  {journal} {\bibinfo
  {journal} {Phys. Rev. B}\ }\textbf {\bibinfo {volume} {87}},\ \bibinfo
  {pages} {214507} (\bibinfo {year} {2013})}\BibitemShut {NoStop}%
\bibitem [{\citenamefont {Madeira}\ \emph {et~al.}(2016)\citenamefont
  {Madeira}, \citenamefont {Vitiello}, \citenamefont {Gandolfi},\ and\
  \citenamefont {Schmidt}}]{mad16}%
  \BibitemOpen
  \bibfield  {author} {\bibinfo {author} {\bibfnamefont {L.}~\bibnamefont
  {Madeira}}, \bibinfo {author} {\bibfnamefont {S.~A.}\ \bibnamefont
  {Vitiello}}, \bibinfo {author} {\bibfnamefont {S.}~\bibnamefont {Gandolfi}},
  \ and\ \bibinfo {author} {\bibfnamefont {K.~E.}\ \bibnamefont {Schmidt}},\
  }\href {\doibase 10.1103/PhysRevA.93.043604} {\bibfield  {journal} {\bibinfo
  {journal} {Phys. Rev. A}\ }\textbf {\bibinfo {volume} {93}},\ \bibinfo
  {pages} {043604} (\bibinfo {year} {2016})}\BibitemShut {NoStop}%
\bibitem [{\citenamefont {Zwierlein}\ \emph {et~al.}(2005)\citenamefont
  {Zwierlein}, \citenamefont {Abo-Shaeer}, \citenamefont {Schirotzek},
  \citenamefont {Schunck},\ and\ \citenamefont {Ketterle}}]{zwi05}%
  \BibitemOpen
  \bibfield  {author} {\bibinfo {author} {\bibfnamefont {M.~W.}\ \bibnamefont
  {Zwierlein}}, \bibinfo {author} {\bibfnamefont {J.~R.}\ \bibnamefont
  {Abo-Shaeer}}, \bibinfo {author} {\bibfnamefont {A.}~\bibnamefont
  {Schirotzek}}, \bibinfo {author} {\bibfnamefont {C.~H.}\ \bibnamefont
  {Schunck}}, \ and\ \bibinfo {author} {\bibfnamefont {W.}~\bibnamefont
  {Ketterle}},\ }\href {\doibase 10.1038/nature03858} {\bibfield  {journal}
  {\bibinfo  {journal} {Nature}\ }\textbf {\bibinfo {volume} {435}},\ \bibinfo
  {pages} {1047} (\bibinfo {year} {2005})}\BibitemShut {NoStop}%
\bibitem [{\citenamefont {Berezinsky}(1971)}]{ber70}%
  \BibitemOpen
  \bibfield  {author} {\bibinfo {author} {\bibfnamefont {V.~L.}\ \bibnamefont
  {Berezinsky}},\ }\href@noop {} {\bibfield  {journal} {\bibinfo  {journal}
  {Sov. Phys. JETP}\ }\textbf {\bibinfo {volume} {32}},\ \bibinfo {pages} {493}
  (\bibinfo {year} {1971})}\BibitemShut {NoStop}%
%%CITATION = SPHJA,32,493;%%
\bibitem [{\citenamefont {Kosterlitz}\ and\ \citenamefont
  {Thouless}(1972)}]{kos72}%
  \BibitemOpen
  \bibfield  {author} {\bibinfo {author} {\bibfnamefont {J.~M.}\ \bibnamefont
  {Kosterlitz}}\ and\ \bibinfo {author} {\bibfnamefont {D.~J.}\ \bibnamefont
  {Thouless}},\ }\href {http://stacks.iop.org/0022-3719/5/i=11/a=002}
  {\bibfield  {journal} {\bibinfo  {journal} {J. Phys. C}\ }\textbf {\bibinfo
  {volume} {5}},\ \bibinfo {pages} {L124} (\bibinfo {year} {1972})}\BibitemShut
  {NoStop}%
\bibitem [{\citenamefont {Botelho}\ and\ \citenamefont {S\'a~de
  Melo}(2006)}]{bot06}%
  \BibitemOpen
  \bibfield  {author} {\bibinfo {author} {\bibfnamefont {S.~S.}\ \bibnamefont
  {Botelho}}\ and\ \bibinfo {author} {\bibfnamefont {C.~A.~R.}\ \bibnamefont
  {S\'a~de Melo}},\ }\href {\doibase 10.1103/PhysRevLett.96.040404} {\bibfield
  {journal} {\bibinfo  {journal} {Phys. Rev. Lett.}\ }\textbf {\bibinfo
  {volume} {96}},\ \bibinfo {pages} {040404} (\bibinfo {year}
  {2006})}\BibitemShut {NoStop}%
\bibitem [{\citenamefont {Carlson}\ \emph {et~al.}(2013)\citenamefont
  {Carlson}, \citenamefont {Gandolfi},\ and\ \citenamefont {Gezerlis}}]{bro13}%
  \BibitemOpen
  \bibfield  {author} {\bibinfo {author} {\bibfnamefont {J.}~\bibnamefont
  {Carlson}}, \bibinfo {author} {\bibfnamefont {S.}~\bibnamefont {Gandolfi}}, \
  and\ \bibinfo {author} {\bibfnamefont {A.}~\bibnamefont {Gezerlis}},\
  }\href@noop {} {\emph {\bibinfo {title} {{F}ifty {Y}ears of {N}uclear
  {BCS}}}},\ edited by\ \bibinfo {editor} {\bibfnamefont {R.~A.}\ \bibnamefont
  {Broglia}}\ and\ \bibinfo {editor} {\bibfnamefont {V.}~\bibnamefont
  {Zelevinsky}}\ (\bibinfo  {publisher} {World Scientific Publishing Company},\
  \bibinfo {year} {2013})\BibitemShut {NoStop}%
\bibitem [{\citenamefont {Gezerlis}\ and\ \citenamefont
  {Carlson}(2008)}]{Gezerlis:2008}%
  \BibitemOpen
  \bibfield  {author} {\bibinfo {author} {\bibfnamefont {A.}~\bibnamefont
  {Gezerlis}}\ and\ \bibinfo {author} {\bibfnamefont {J.}~\bibnamefont
  {Carlson}},\ }\href@noop {} {\bibfield  {journal} {\bibinfo  {journal} {Phys.
  Rev. C}\ }\textbf {\bibinfo {volume} {77}},\ \bibinfo {pages} {032801(R)}
  (\bibinfo {year} {2008})}\BibitemShut {NoStop}%
\bibitem [{\citenamefont {Gezerlis}\ and\ \citenamefont
  {Carlson}(2010)}]{gez10}%
  \BibitemOpen
  \bibfield  {author} {\bibinfo {author} {\bibfnamefont {A.}~\bibnamefont
  {Gezerlis}}\ and\ \bibinfo {author} {\bibfnamefont {J.}~\bibnamefont
  {Carlson}},\ }\href {\doibase 10.1103/PhysRevC.81.025803} {\bibfield
  {journal} {\bibinfo  {journal} {Phys. Rev. C}\ }\textbf {\bibinfo {volume}
  {81}},\ \bibinfo {pages} {025803} (\bibinfo {year} {2010})}\BibitemShut
  {NoStop}%
\bibitem [{\citenamefont {De~Blasio}\ and\ \citenamefont
  {Elgar\o{}y}(1999)}]{bla99}%
  \BibitemOpen
  \bibfield  {author} {\bibinfo {author} {\bibfnamefont {F.~V.}\ \bibnamefont
  {De~Blasio}}\ and\ \bibinfo {author} {\bibfnamefont {O.}~\bibnamefont
  {Elgar\o{}y}},\ }\href {\doibase 10.1103/PhysRevLett.82.1815} {\bibfield
  {journal} {\bibinfo  {journal} {Phys. Rev. Lett.}\ }\textbf {\bibinfo
  {volume} {82}},\ \bibinfo {pages} {1815} (\bibinfo {year}
  {1999})}\BibitemShut {NoStop}%
\bibitem [{\citenamefont {Yu}\ and\ \citenamefont {Bulgac}(2003)}]{yu03}%
  \BibitemOpen
  \bibfield  {author} {\bibinfo {author} {\bibfnamefont {Y.}~\bibnamefont
  {Yu}}\ and\ \bibinfo {author} {\bibfnamefont {A.}~\bibnamefont {Bulgac}},\
  }\href {\doibase 10.1103/PhysRevLett.90.161101} {\bibfield  {journal}
  {\bibinfo  {journal} {Phys. Rev. Lett.}\ }\textbf {\bibinfo {volume} {90}},\
  \bibinfo {pages} {161101} (\bibinfo {year} {2003})}\BibitemShut {NoStop}%
\bibitem [{\citenamefont {Ravenhall}\ \emph {et~al.}(1983)\citenamefont
  {Ravenhall}, \citenamefont {Pethick},\ and\ \citenamefont {Wilson}}]{rav83}%
  \BibitemOpen
  \bibfield  {author} {\bibinfo {author} {\bibfnamefont {D.~G.}\ \bibnamefont
  {Ravenhall}}, \bibinfo {author} {\bibfnamefont {C.~J.}\ \bibnamefont
  {Pethick}}, \ and\ \bibinfo {author} {\bibfnamefont {J.~R.}\ \bibnamefont
  {Wilson}},\ }\href {\doibase 10.1103/PhysRevLett.50.2066} {\bibfield
  {journal} {\bibinfo  {journal} {Phys. Rev. Lett.}\ }\textbf {\bibinfo
  {volume} {50}},\ \bibinfo {pages} {2066} (\bibinfo {year}
  {1983})}\BibitemShut {NoStop}%
\bibitem [{\citenamefont {Sadd}\ \emph {et~al.}(1997)\citenamefont {Sadd},
  \citenamefont {Chester},\ and\ \citenamefont {Reatto}}]{sad97}%
  \BibitemOpen
  \bibfield  {author} {\bibinfo {author} {\bibfnamefont {M.}~\bibnamefont
  {Sadd}}, \bibinfo {author} {\bibfnamefont {G.~V.}\ \bibnamefont {Chester}}, \
  and\ \bibinfo {author} {\bibfnamefont {L.}~\bibnamefont {Reatto}},\ }\href
  {\doibase 10.1103/PhysRevLett.79.2490} {\bibfield  {journal} {\bibinfo
  {journal} {Phys. Rev. Lett.}\ }\textbf {\bibinfo {volume} {79}},\ \bibinfo
  {pages} {2490} (\bibinfo {year} {1997})}\BibitemShut {NoStop}%
\bibitem [{\citenamefont {Ortiz}\ and\ \citenamefont
  {Ceperley}(1995)}]{Ortiz:1995}%
  \BibitemOpen
  \bibfield  {author} {\bibinfo {author} {\bibfnamefont {G.}~\bibnamefont
  {Ortiz}}\ and\ \bibinfo {author} {\bibfnamefont {D.~M.}\ \bibnamefont
  {Ceperley}},\ }\href {\doibase 10.1103/PhysRevLett.75.4642} {\bibfield
  {journal} {\bibinfo  {journal} {Phys. Rev. Lett.}\ }\textbf {\bibinfo
  {volume} {75}},\ \bibinfo {pages} {4642} (\bibinfo {year}
  {1995})}\BibitemShut {NoStop}%
\bibitem [{\citenamefont {Khuri}\ \emph {et~al.}(2009)\citenamefont {Khuri},
  \citenamefont {Martin}, \citenamefont {Richard},\ and\ \citenamefont
  {Wu}}]{khu09}%
  \BibitemOpen
  \bibfield  {author} {\bibinfo {author} {\bibfnamefont {N.~N.}\ \bibnamefont
  {Khuri}}, \bibinfo {author} {\bibfnamefont {A.}~\bibnamefont {Martin}},
  \bibinfo {author} {\bibfnamefont {J.-M.}\ \bibnamefont {Richard}}, \ and\
  \bibinfo {author} {\bibfnamefont {T.~T.}\ \bibnamefont {Wu}},\ }\href
  {\doibase http://dx.doi.org/10.1063/1.3167803} {\bibfield  {journal}
  {\bibinfo  {journal} {Journal of Mathematical Physics}\ }\textbf {\bibinfo
  {volume} {50}},\ \bibinfo {eid} {072105} (\bibinfo {year} {2009}),\
  http://dx.doi.org/10.1063/1.3167803}\BibitemShut {NoStop}%
\bibitem [{\citenamefont {Adhikari}\ \emph {et~al.}(1986)\citenamefont
  {Adhikari}, \citenamefont {Gibson},\ and\ \citenamefont {Lim}}]{adh86}%
  \BibitemOpen
  \bibfield  {author} {\bibinfo {author} {\bibfnamefont {S.~K.}\ \bibnamefont
  {Adhikari}}, \bibinfo {author} {\bibfnamefont {W.~G.}\ \bibnamefont
  {Gibson}}, \ and\ \bibinfo {author} {\bibfnamefont {T.~K.}\ \bibnamefont
  {Lim}},\ }\href {\doibase http://dx.doi.org/10.1063/1.451572} {\bibfield
  {journal} {\bibinfo  {journal} {The Journal of Chemical Physics}\ }\textbf
  {\bibinfo {volume} {85}},\ \bibinfo {pages} {5580} (\bibinfo {year}
  {1986})}\BibitemShut {NoStop}%
\bibitem [{\citenamefont {Carlson}\ \emph {et~al.}(2003)\citenamefont
  {Carlson}, \citenamefont {Chang}, \citenamefont {Pandharipande},\ and\
  \citenamefont {Schmidt}}]{car03}%
  \BibitemOpen
  \bibfield  {author} {\bibinfo {author} {\bibfnamefont {J.}~\bibnamefont
  {Carlson}}, \bibinfo {author} {\bibfnamefont {S.-Y.}\ \bibnamefont {Chang}},
  \bibinfo {author} {\bibfnamefont {V.~R.}\ \bibnamefont {Pandharipande}}, \
  and\ \bibinfo {author} {\bibfnamefont {K.~E.}\ \bibnamefont {Schmidt}},\
  }\href {\doibase 10.1103/PhysRevLett.91.050401} {\bibfield  {journal}
  {\bibinfo  {journal} {Phys. Rev. Lett.}\ }\textbf {\bibinfo {volume} {91}},\
  \bibinfo {pages} {050401} (\bibinfo {year} {2003})}\BibitemShut {NoStop}%
\bibitem [{\citenamefont {Gezerlis}\ \emph {et~al.}(2009)\citenamefont
  {Gezerlis}, \citenamefont {Gandolfi}, \citenamefont {Schmidt},\ and\
  \citenamefont {Carlson}}]{gez09}%
  \BibitemOpen
  \bibfield  {author} {\bibinfo {author} {\bibfnamefont {A.}~\bibnamefont
  {Gezerlis}}, \bibinfo {author} {\bibfnamefont {S.}~\bibnamefont {Gandolfi}},
  \bibinfo {author} {\bibfnamefont {K.~E.}\ \bibnamefont {Schmidt}}, \ and\
  \bibinfo {author} {\bibfnamefont {J.}~\bibnamefont {Carlson}},\ }\href
  {\doibase 10.1103/PhysRevLett.103.060403} {\bibfield  {journal} {\bibinfo
  {journal} {Phys. Rev. Lett.}\ }\textbf {\bibinfo {volume} {103}},\ \bibinfo
  {pages} {060403} (\bibinfo {year} {2009})}\BibitemShut {NoStop}%
\bibitem [{\citenamefont {Gandolfi}(2014)}]{gan14}%
  \BibitemOpen
  \bibfield  {author} {\bibinfo {author} {\bibfnamefont {S.}~\bibnamefont
  {Gandolfi}},\ }\href {http://stacks.iop.org/1742-6596/529/i=1/a=012011}
  {\bibfield  {journal} {\bibinfo  {journal} {Journal of Physics: Conference
  Series}\ }\textbf {\bibinfo {volume} {529}},\ \bibinfo {pages} {012011}
  (\bibinfo {year} {2014})}\BibitemShut {NoStop}%
\bibitem [{\citenamefont {{Bouchaud, J.P.}}\ \emph {et~al.}(1988)\citenamefont
  {{Bouchaud, J.P.}}, \citenamefont {{Georges, A.}},\ and\ \citenamefont
  {{Lhuillier, C.}}}]{bou88}%
  \BibitemOpen
  \bibfield  {author} {\bibinfo {author} {\bibnamefont {{Bouchaud, J.P.}}},
  \bibinfo {author} {\bibnamefont {{Georges, A.}}}, \ and\ \bibinfo {author}
  {\bibnamefont {{Lhuillier, C.}}},\ }\href {\doibase
  10.1051/jphys:01988004904055300} {\bibfield  {journal} {\bibinfo  {journal}
  {J. Phys. France}\ }\textbf {\bibinfo {volume} {49}},\ \bibinfo {pages} {553}
  (\bibinfo {year} {1988})}\BibitemShut {NoStop}%
\bibitem [{\citenamefont {Gandolfi}\ \emph {et~al.}(2009)\citenamefont
  {Gandolfi}, \citenamefont {Illarionov}, \citenamefont {Pederiva},
  \citenamefont {Schmidt},\ and\ \citenamefont {Fantoni}}]{gan09}%
  \BibitemOpen
  \bibfield  {author} {\bibinfo {author} {\bibfnamefont {S.}~\bibnamefont
  {Gandolfi}}, \bibinfo {author} {\bibfnamefont {A.~Y.}\ \bibnamefont
  {Illarionov}}, \bibinfo {author} {\bibfnamefont {F.}~\bibnamefont
  {Pederiva}}, \bibinfo {author} {\bibfnamefont {K.~E.}\ \bibnamefont
  {Schmidt}}, \ and\ \bibinfo {author} {\bibfnamefont {S.}~\bibnamefont
  {Fantoni}},\ }\href {\doibase 10.1103/PhysRevC.80.045802} {\bibfield
  {journal} {\bibinfo  {journal} {Phys. Rev. C}\ }\textbf {\bibinfo {volume}
  {80}},\ \bibinfo {pages} {045802} (\bibinfo {year} {2009})}\BibitemShut
  {NoStop}%
\bibitem [{\citenamefont {Foulkes}\ \emph {et~al.}(2001)\citenamefont
  {Foulkes}, \citenamefont {Mitas}, \citenamefont {Needs},\ and\ \citenamefont
  {Rajagopal}}]{fou01}%
  \BibitemOpen
  \bibfield  {author} {\bibinfo {author} {\bibfnamefont {W.~M.~C.}\
  \bibnamefont {Foulkes}}, \bibinfo {author} {\bibfnamefont {L.}~\bibnamefont
  {Mitas}}, \bibinfo {author} {\bibfnamefont {R.~J.}\ \bibnamefont {Needs}}, \
  and\ \bibinfo {author} {\bibfnamefont {G.}~\bibnamefont {Rajagopal}},\ }\href
  {\doibase 10.1103/RevModPhys.73.33} {\bibfield  {journal} {\bibinfo
  {journal} {Rev. Mod. Phys.}\ }\textbf {\bibinfo {volume} {73}},\ \bibinfo
  {pages} {33} (\bibinfo {year} {2001})}\BibitemShut {NoStop}%
\bibitem [{Note1()}]{Note1}%
  \BibitemOpen
  \bibinfo {note} {See Supplemental Material at [\protect \textit {url to be
  inserted by publisher}] for the variational parameters of Eqs.~(\ref
  {eq:bulk}), (\ref {eq:disk}), and (\ref {eq:vortex})}\BibitemShut {NoStop}%
\bibitem [{\citenamefont {Casula}\ \emph {et~al.}(2004)\citenamefont {Casula},
  \citenamefont {Attaccalite},\ and\ \citenamefont {Sorella}}]{cas04}%
  \BibitemOpen
  \bibfield  {author} {\bibinfo {author} {\bibfnamefont {M.}~\bibnamefont
  {Casula}}, \bibinfo {author} {\bibfnamefont {C.}~\bibnamefont {Attaccalite}},
  \ and\ \bibinfo {author} {\bibfnamefont {S.}~\bibnamefont {Sorella}},\ }\href
  {\doibase http://dx.doi.org/10.1063/1.1794632} {\bibfield  {journal}
  {\bibinfo  {journal} {The Journal of Chemical Physics}\ }\textbf {\bibinfo
  {volume} {121}},\ \bibinfo {pages} {7110} (\bibinfo {year}
  {2004})}\BibitemShut {NoStop}%
\bibitem [{\citenamefont {Ceperley}\ and\ \citenamefont {Kalos}(1986)}]{cep86}%
  \BibitemOpen
  \bibfield  {author} {\bibinfo {author} {\bibfnamefont {D.}~\bibnamefont
  {Ceperley}}\ and\ \bibinfo {author} {\bibfnamefont {H.}~\bibnamefont
  {Kalos}},\ }\href@noop {} {\emph {\bibinfo {title} {{M}onte {C}arlo {M}ethods
  in {S}tatistics {P}hysics {Q}uantum {M}any-{B}ody {P}roblems}}},\ edited by\
  \bibinfo {editor} {\bibfnamefont {K.~E.}\ \bibnamefont {Binder}},\
  Vol.~\bibinfo {volume} {7}\ (\bibinfo  {publisher} {Springer-Verlag},\
  \bibinfo {year} {1986})\BibitemShut {NoStop}%
\bibitem [{\citenamefont {K\"onig}\ \emph {et~al.}(2004)\citenamefont
  {K\"onig}, \citenamefont {Roth},\ and\ \citenamefont {Mecke}}]{kon04}%
  \BibitemOpen
  \bibfield  {author} {\bibinfo {author} {\bibfnamefont {P.-M.}\ \bibnamefont
  {K\"onig}}, \bibinfo {author} {\bibfnamefont {R.}~\bibnamefont {Roth}}, \
  and\ \bibinfo {author} {\bibfnamefont {K.~R.}\ \bibnamefont {Mecke}},\ }\href
  {\doibase 10.1103/PhysRevLett.93.160601} {\bibfield  {journal} {\bibinfo
  {journal} {Phys. Rev. Lett.}\ }\textbf {\bibinfo {volume} {93}},\ \bibinfo
  {pages} {160601} (\bibinfo {year} {2004})}\BibitemShut {NoStop}%
\bibitem [{Note2()}]{Note2}%
  \BibitemOpen
  \bibinfo {note} {After our calculations were done, it was pointed out to us
  that pseudogap phenomena occurring in 2D and 3D Fermi gases can be related in
  a universal way through a variable that spans the BEC-BCS crossover \cite
  {mar15}. Further studies are necessary to determine if this universality
  holds for other quantities, such as the density and the probability current
  density per particle. This would provide a very clean way of comparing 2D and
  3D results.}\BibitemShut {Stop}%
\bibitem [{\citenamefont {Marsiglio}\ \emph {et~al.}(2015)\citenamefont
  {Marsiglio}, \citenamefont {Pieri}, \citenamefont {Perali}, \citenamefont
  {Palestini},\ and\ \citenamefont {Strinati}}]{mar15}%
  \BibitemOpen
  \bibfield  {author} {\bibinfo {author} {\bibfnamefont {F.}~\bibnamefont
  {Marsiglio}}, \bibinfo {author} {\bibfnamefont {P.}~\bibnamefont {Pieri}},
  \bibinfo {author} {\bibfnamefont {A.}~\bibnamefont {Perali}}, \bibinfo
  {author} {\bibfnamefont {F.}~\bibnamefont {Palestini}}, \ and\ \bibinfo
  {author} {\bibfnamefont {G.~C.}\ \bibnamefont {Strinati}},\ }\href {\doibase
  10.1103/PhysRevB.91.054509} {\bibfield  {journal} {\bibinfo  {journal} {Phys.
  Rev. B}\ }\textbf {\bibinfo {volume} {91}},\ \bibinfo {pages} {054509}
  (\bibinfo {year} {2015})}\BibitemShut {NoStop}%
\end{thebibliography}%
\end{document}